\documentclass[a4paper,oneside,Palatino]{article}
\usepackage{latexsym}
\usepackage[dvips]{color}
\usepackage[dvips]{graphicx}
\usepackage[dvips]{graphics}
\usepackage{amsmath}
\usepackage{amssymb}
\usepackage{latexsym}
\usepackage{cite}
\usepackage[dvips]{color}
\usepackage[dvips]{graphicx} 

\usepackage{appendix}
   \usepackage{caption}

\usepackage{amsmath} 
\catcode`@=11
\renewcommand{\theequation}{\thesection.\arabic{equation}}
\@addtoreset{equation}{section}
\catcode`@=12

\title{\centerline{\small BHU/HEP/2020 \hfill hep-ph/2002276}\bigskip
\bf Exploring scalar-photon interactions in energetic astrophysical events.}

\author{Ankur Chaubey$^a$, Manoj K. Jaiswal$^a$ and Avijit K. Ganguly$^a$\thanks{e-mail addresses: avijitk@hotmail.com }  \thanks{corresponding author},\\ 
\normalsize 
$a)$ Institute of Science, Department of Physics, Banaras Hindu University, 
Varanasi- 221005, INDIA.
\\
\normalsize 
}

\begin{document}
\maketitle
\begin{abstract} 
Scalar fields like dilaton appear in quantum field theory (QFT) due to scale symmetry
breaking. Their appeal also extends to modified theories of gravity, like 
$F(R)$ gravity, Horva Lifshitz gravity etc.  In unified theories they make their  
appearance through compactification of the extra dimension. 
Apart from resolving the issues of compactification scale and size, 
the particles of their fields can also turn out to be excellent 
candidate to solve the dark energy (DE)  and dark matter (DM) problem of the universe. 
In this work we  study their mixing dynamics with photons in a magnetized 
media,  by incorporating  the effect of parity violating part of the photon 
polarization  tensor, evaluated in a finite density magnetized media.
This piece, though in general is  odd in the external magnetic field 
strength $eB$; in this work we however have retained terms to $O$($eB$).
We are able to demonstrate in this work that, in magnetized medium 
a dilatonic scalar field $(\phi)$  can excite the two transverse degrees 
of freedom (DOF) of the photons. One due to direct coupling and the other 
indirectly through the parity violating term originating due to magnetized 
medium effects. This results in the mixing dynamics being governed by, 
$3\times 3$ mixing  matrices. This mixing results in making the underlying 
media optically active. In this work we focus on the spectro-polarimetric 
imprints of these particles, on the spectra of the electromagnetic (EM) 
fields  of Gamma Ray Bursters (GRB). Focusing on a range of parameters 
(i.e., magnetic field strength, plasma frequency $(\omega_{p})$, size of 
the magnetized volume, coupling strength to photons and their mass) 
we make an attempt to point out how space-borne detectors should be 
designed to optimise their detection possibility.
\end{abstract}

\section{Introduction}
The study of scale symmetry and it's consequences on the dynamics of particles
has drawn the attention for some time now.
The particles appearing  as Goldstone bosons of a 
spontaneously broken scale symmetry (termed dilaton, $\phi(x)$) \cite{Kim, donoghue},
 have emerged from studies in QFT.
There are many theories those predict the existence of dilatons. 
Apart from QFT, they appear in higher dimensional unified theories, 
for instance, in five dimensional Kaluza-Klein theory, they appear
 as the five-five component of the five dimensional metric, formulated 
to unify gravity with electromagnetism.  In string and super-string theory
they appear from compactification of the extra dimensions and are called 
string dilaton or modouli \cite{gasperini, bobby_Maharana}.\\

\indent
In some scale invariant extensions of standard model, they are made to
communicate with the standard model sector via an underlying conformal 
sector, where they acquire mass due to breaking of the conformal 
in-variance\cite{coriano,salam}. This physics of these models are phenomenologically 
rich with predicting power that can be tested in collider based experiments. 
On the other hand dilatons of unified theories, acquire their masses from 
the  curvature of the extra dimension.\\
\indent
They couple to the standard model fields by the trace of their energy 
momentum tensor $T^{\mu}_{\nu}$, associated with the anomalous divergence 
of the dilaton 4-current\footnote{The nonzero anomalous divergence, even 
for mass-less particles may realized due to scaling violation through radiative 
corrections}. Due to this, dilatons may induce other observable 
signatures, like dilatonic fifth force:\cite{fifthforce1,fifthforce2,
fifthforce3,fifthforce4,booby}; bending of light \cite{bending}, 
violation of equivalence principle \cite{equiv}, decay into two photons
and optical activity  \cite{ Miani, Raffelt, Kahniashvili, Ganguly-parthasarathy, Giannotti2}
 in external magnetic field ($B$).  
When the last two phenomena -- that is decay  of $\phi$ into two mass-less spin one photons
and optical activity--follow from the interaction Lagrangian,\\
\begin{eqnarray}
L_{int}=  -\frac{1}{4M}\phi F^{\mu\nu}F_{\mu\nu}.
\label{lint}
\end{eqnarray}
\noindent
In equation (\ref {lint}) $F^{\mu\nu}$  is the usual field strength 
tensor for EM field. And $ M $ is the symmetry breaking scale
related to the inverse of the coupling constant $g_{\phi\gamma\gamma}$, 
between the quanta of  scalar $(\phi)$  and photon $(\gamma)$ fields. Equation 
(\ref{lint}) leads to their lifetime $\tau_{\phi}$ against 
decay to two photons $\phi \to \gamma \gamma$, given by
$\tau_{\phi}  \sim \frac{1}{g^2_{\phi\gamma\gamma}m_{\phi}^3} $ \cite{Kim}.
If the life time of these particles for some values of $g_{\phi\gamma\gamma}$
and $m_{\phi}$, turn out to be comparable to the age of the universe, 
then the particles of the field $\phi$ will turn out to be excellent 
candidates for DM. Thus, simultaneously solving the two out standing 
problems of contemporary physics. There are other particles those  produce
 similar signals can be found in
  \cite{shahbad, nieves, servant, axion-fifthforce, diff-axion-scalar-lab-tests, Parthamajumdar, RaffeltChi, Ringwald,     Giannotti1 , ayla, yanagida, Giannotti, palash-prima, Mclerran, Cast, Castnature, Armengaud, xenon, MADMAX, DARWIN, Irastorza, Millar, Kartavtsev, subir, Masaki, Ganguly-jaiswal, JKPS, manoj  } but we will not  discuss those  in this work. \\ 
\indent
Given the state of our current understanding that, about  27 $\%$ of the total 
matter-energy density is in the form of DM, it is possible
to find what percentage of the total DM density is composed of $\phi$, at some
 epoch $t$, as cosmological  relic density  $\rho(t)$, from \cite{kolb-turner}:
\begin{eqnarray}
\rho(t) = \rho_{d} \left(\frac{\eta_d}{\eta}\right)^3 e^{-\frac{t}{\tau_{\phi}}}  
= \frac{\zeta(3)}{\pi^2}\frac{g^{*}(t)}{g_{*d}}T^3(t) 
e^{-\frac{t}{\tau_{\phi}}}.
\label{relic-density}
\end{eqnarray}

\indent
In equation (\ref{relic-density})  $\rho_{d}$ corresponds to 
the density and $\eta_d$ the magnitude of the scale factor of Friedman-Robertson-Walker metric. 
And $g_{*d}$ along with $g^{*}(t)$ are the number of DOF available at the time of 
decoupling and the same at the epoch t, respectively. The last line in 
equation (\ref{relic-density}) has been obtained demanding entropy conservation 
in the co-moving volume of the universe, from the time of decoupling to the epoch $t$.
 Since the same depends crucially on the life-time $\tau_{\phi}$, 
that in-turn depends on coupling constant $g_{\phi\gamma\gamma}=\frac{1}{4M}$ and 
the scalar  mass $m_{\phi}$ -- therefore the estimations of them are of utmost 
importance. In this study we focus on their estimation from  the EM signals originating
 due to the energetic activities taking place
in far away magnetized astrophysical objects. To that end, we have taken a
 spectro-polarimetric route in this paper, 
to estimate the parameters (mass and coupling constant) associated  with these 
particles (dilatonic scalars) by studying  their mixing  dynamics in 
presence of magnetized plasma present in the  GRB environments. 
 We also indicate how  such (spectro-polarimetric) analysis can be used to 
design the space-borne gamma-ray or X-ray detectors -- optimally -- 
to detect dilaton signatures through EM signals coming from GRB .\\

 At this juncture  we would like to digress a little, so as to pay attention to other possible physical sources those may 
contribute to the polarization of electromagnetic field, coming from far away sources. One of the possible sources that can contribute the polarization of electromagnetic beam is contribution from magnetized medium and the other can be coming due to the presence of pseudoscalar particles like axions or majorons, those couple to photons through mass dimension-5 operators
 Rotation of the plane of polarization of light due to magnetized medium, that is referred usually in literature as Faraday effect, takes place
 in a material medium having nonzero chemical potential, and a magnetic field $B$ when the magnetic field is oriented along the direction of propagation of the photon $k$. On the other hand the effect due to the scalar dilaton or pseudoscalar field, takes place, when the component of the magnetic field,  is perpendicular to the wave vector $k$. So these two effects can be distinguished from each other by making the external magnetic field parallel or perpendicular to the propagation direction of the photons. It is also worth noting that, rate of rotation for plane of polarization for Faraday effect is inversely proportional to the square of the energy $\omega$ of the photons  of light. Hence the same can also  serve as a distinct feature to identify magnetized matter induced polarization effect from dilatons or axions.  Moreover the degree of circular polarization associated with the beam of light passing through magnetized medium turns out to be zero. The details of these can be found in the Appendix {\bf A}. \\

\indent
Now coming to the issue of  distinguishing  scalars(dilatons) from pseudoscalar axions, it should be noted that in magnetized vacuum, dilaton mixes with polarized light having plane of polarization oriented along the magnetic field and axion mixes with light having plane of polarization orthogonal to the magnetic field. This simple picture however gets complicated with the incorporation of magnetized matter effects. We  will come back to this issue in a separate publication.  Having taken these extra-dilaton sources, contributing to polarization of light from astrophysical sources, we turn our attention to the investigations carried out in this work.

\indent
During the course of this investigation, we have achieved few new things. They 
include the following (i) a better understanding of the unique nature of the
set of basis vectors and the form-factors, those appear in the description  
of the gauge fields (i.e., of the photons), for a system in an external 
magnetic field $eB$, and magnetized media. We provide the transformation 
properties of the EM form-factors and other factors under Charge conjugation {\bf C}, Parity {\bf P} and 
Time reversal {\bf T} and use them to justify the coupling between the different DOF 
available to the system, leading to a 3 $\times$ 3 mixing matrix; 
(ii) We next  provide the analytical route to diagonalise this 
matrix exactly, using an unitary similarity transformation. (iii) The resulting
equations of motion obtained thereby are also exact. (iv) The numerical 
estimates  of the Stokes parameters obtained from the numerical estimates 
of the form-factors, thus are also without any approximations.\\

\indent
The organisation of this document is as follows: In section two we introduce the
details, about properties of the gauge fields (GF), the form-factors those 
describe them (GF) and their transformation properties under {\bf C}, {\bf P} 
and  {\bf T}. This is followed by the description of the photon polarization 
tensor in a magnetized medium in section three, called inclusion of matter effects. In section four we move on to analysing the 
equations of motion for $(\gamma-\phi)$ interacting system in a magnetized 
medium and demonstrate that  the mixing matrix for $\phi F_{\mu\nu}F^{\mu\nu}$ 
interaction, turns out to be $3\times 3$ instead of $2\times 2$, that is usually encountered in magnetized vacuum, 
or unmagnetized plasma. We discuss
the exact analytic diagonalization of the same in the following subsection.  
Subsequently we justify the same from discrete symmetry point of view.
 The solutions of the field equations followed by construction
of the Stokes parameters is obtained in section five.
In section six we introduce  a typical GRB model and its environment that being 
used in this analysis and the polarization signals one would get from the same, for some bench mark values of $g_{\phi\gamma\gamma}$ and $m_{\phi}$, the geometry of 
the GRB fireball and plasma frequency. The  possible EM signatures of dilaton 
interaction  from such environments is presented in section seven. Section 
 eight houses a discussion on the relevance of our analysis to space-borne 
 detectors. In  section nine, we conclude by providing an outlook for possible  future directions  of investigation. And lastly, we have provided  an  appendix  that deals with the details of polarization evaluation in a magnetized media. Few important details regarding  discrete symmetry transformations and their effects on equations of motion can be found in the supplementary materials in a separate work, titled:  "Supplementary materials for  Exploring scalar-photon interactions in energetic astrophysical events."

\section{Electromagnetic form-factors for $A^{\nu}(k)$}
\indent 
In the standard formulation of  the massless abelian gauge theory, that
 describes the dynamics of photons, the action is written as,
\begin{eqnarray}
S = -\frac{1}{4}\int F^{\mu\nu}F_{\mu\nu} d^4x 
\label{action}
\end{eqnarray}
where in equation (\ref{action}),  the field strength tensor, $F^{\mu\nu}= \partial^{\mu} A^{\nu}(x) -  \partial^{\nu} A^{\mu}(x)$,  and $A^{\nu}(x)$
defines the gauge potentials having four DOF. The dynamics 
of these fields in vacuum are described by two transverse $\pm 1$ helicity 
states. \\  
\indent
In contrast to vacuum, photons in a medium, acquire one additional DOF,
(the third) longitudinal DOF-- in addition to the two (existing)
transverse degrees of freedom. In a situation like this, if there 
exists an external magnetic field $B$ too,  then the gauge fields 
$A^{\nu}(k)$ corresponding to the in medium 
photons, can be expressed (in momentum space), in terms of four EM 
form-factors: $  A_{\parallel}(k)$, $  A_{\perp}(k)  $, $ A_{L}(k) $, $A_{gf}(k)$ and four orthonormal four vectors $(b^{(1)\nu}, I^{\nu}, {\tilde{u}}^{\nu}, k^{\nu} )$ constructed 
out of the available 4-vectors and tensors for the system  (in hand). They are given by :
\begin{eqnarray}
A^{\nu}(k)= A_{\parallel}(k){N}_1 b^{(1)\nu}+ A_{\perp}(k) {N}_2I^{\nu}+ A_{L}(k)
{N}_L\tilde{u}^{\nu} + {N}_k A_{gf}(k)k^{\nu}.
\label{gauge-pot2}
\end{eqnarray}

\noindent
Here,  $N_i$s are the normalisation constants. Rewriting  $A^{\nu}(k)$ in terms of unit vectors;  $\hat{b}^{(1)\nu} = N_{1}{b}^{(1)\nu}$, $ \hat{I}^{\nu} = N_{2}{I}^{\nu}$,  $ \hat{\tilde{u}}^{\nu} = N_{L}{b}^{(1)\nu}$  and $\hat{k}^{\nu} = N_{k}k^{\nu}$ we can rewrite equation (\ref{gauge-pot2}) in the following form,

\begin{eqnarray}
A^{\nu}(k)= A_{\parallel}(k) \hat{b}^{(1)\nu}+ A_{\perp}(k) \hat{I}^{\nu}+ A_{L}(k)
\hat{\tilde{u}}^{\nu} +  A_{gf}(k)\hat{k}^{\nu}.
\label{gauge-pot3}
\end{eqnarray}
\noindent
The vectors, introduced in equation (\ref{gauge-pot3}), are defined as,

\begin{eqnarray}
 \hat{b}^{(1)\nu} \!\!\! &=& \!\!\! N_{1}k_\mu\bar{F}^{\mu\nu}, ~\hat{I}^{\nu} = N_{2}\left(b^{(2)\nu} - \frac{(\tilde{u}^\mu b^{(2)}_\mu)}{\tilde{u}^2}\tilde{u}^\nu \right), ~
 \hat{{\tilde{u}}}^{\nu} =N_{L}  \left( g^{\mu\nu} - \frac{k^\mu k^\nu}{k^2}\right)u_\mu ,  \nonumber \\
       b^{(2)\nu}& =& k_\mu \tilde{\bar{F}}^{\mu\nu}, ~  \tilde{\bar{F}}^{\mu\nu}= \frac{1}{2}\epsilon^{\mu\nu\lambda\rho}\bar{F}_{\lambda\rho}.
\label{ortho-vectors}
\end{eqnarray}
The normalisation constants, $N_{1}$, $N_{2}$, $N_{L}$ and $N_{k}$ in equation 
(\ref{gauge-pot2}) are given by,
\begin{eqnarray}
N_{1} = \frac{1}{\sqrt{-b^{(1)}_\mu b^{(1)\mu}}}  
=\frac{1}{K_{\perp}B},~
N_{2} = \frac{1}{\sqrt{-I_\mu I^\mu}}=\frac{K}{\omega K_{\perp}B},  
~N_{L} = \frac{1}{\sqrt{-\tilde{u}_\mu \tilde{u}^\mu}}=\frac{k^{2}}{|K|},
\mbox{~and~}
N_k = \frac{1}{\sqrt{-k^2}}
\label{normalizators}
\end{eqnarray}
where $ K_{\perp} = (k^2_1 + k^2_2)^{\frac{1}{2}}$. \\

\subsection{ Degrees of freedom}

\noindent
In field theory, medium effects are incorporated into a system by adding 
a self energy corrected effective Lagrangian to the tree level Lagrangian. 
For electromagnetic theory, this term, in momentum space has the form 
$A^{\mu}(-k) \Pi_{\mu\nu}(k,T,\mu)A^{\nu}(k)$; when, $\Pi_{\mu\nu}(k)$ , the 
polarization tensor, can be expressed in terms of  transverse and longitudinal
form factors, $\Pi_{T}(k,T,\mu)$ and $\Pi_{L}(k,T,\mu)$ as,

\begin{eqnarray}
\Pi_{\mu\nu}(k,T,\mu)= \Pi_{T}(k,T,\mu) \Big[R_{\mu\nu} -Q_{\mu\nu}  \Big]
+ \Pi_{L}(k,T,\mu) Q_{\mu\nu}.
\label{Pol}
\end{eqnarray}

These form factors  happen to be functions of finite temperature (T), finite
chemical potential ($\mu$) and scalars made  out of 
photon four vector $k^{\mu}$ and  centre of mass four velocity of the medium 
$u^{\mu}$ individually,  or  as a combination such as $(k.u)$.
Tensors $R_{\mu\nu}$ and $Q_{\mu\nu}$ are the transverse and longitudinal
projection operators, constructed using the momentum and centre of mass velocity
four vector $u^{\mu}$ as,
\begin{eqnarray}
R_{\mu\nu} = {\tilde{g}_{\mu\nu}}  \mbox{~,~}  
Q_{\mu\nu}=\frac{ {\tilde{u}_{\mu}} {\tilde{u}_{\nu}} }
{ {\tilde{u}_{\mu}} {\tilde{u}^{\mu}} }  \mbox{~and~} 
{\tilde{g}_{\mu\nu}} = g_{\mu\nu} -\frac{k_{\mu} k_{\nu}}{k^2}.
\end{eqnarray}

The transverse and longitudinal form factors have the property that,
in the limit $ \omega =0$ and  $k\to 0$,  $\Pi_{T}$ turns out to be zero and $\Pi_{L}
\to \omega^2_p $ \cite{linde, Fradkin, pal-david, bellac}. The  limit $\omega= 0$ and $k\to 0$, also termed "the" infrared limit, remains  an interesting one  to study the
long wavelength paradigm of the gauge field excitations in a finite density medium. One can analyse infrared    dynamics of the 
system in configuration space by first taking the limit mentioned before,  followed by replacing $ k^{i}  \to  
i \partial^{i} $ in eqn.(\ref{Pol}).  The addition of the background medium  induced pieces  to the effective Lagrangian  may 
not change  the number of  degrees of freedom of the system always. For instance,
 in presence of a background  external electro-magnetic field, the number of physical
degrees of freedom  of quantum corrected U(1) gauge theory with fermions, remains the same as that of the free theory. 
However, the same may not be true when the medium induced quantum corrections are incorporated. The presence of unphysical degrees of freedom in a dynamical system can be inferred from the hessian matrix of the same.
 If the  hessian matrix, 
 
 \begin{eqnarray}
\frac{\partial^2 L _{eff} }
{{\partial{\dot{A_{\mu}}}}{\partial{\dot{A_{\nu}}}}},
\label{HM}
\end{eqnarray} 

\noindent
is noninvertible, the system is constrained,  i.e., the number of dynamical variables are  more than the number of physical degrees of freedom present in the system. In that case, to analyse the dynamics of the system one needs to follow the procedures out lined in \cite{Hanson, Dirac, ECG-MUKUNDA, Maharana}. With finite density effects, incorporated (in the effective Lagrangian) this method, may get very complicated
due to presence of higher derivative terms. However, once the procedures (of constraints analysis) are completed, one can find the number of physical degrees of freedom ($N$)
from the equation,

\begin{eqnarray}
N= \frac {N_{psv} - 2 \times n_1 - n_2}{2}
\end{eqnarray}
where, $N_{psv}$ stands for number of phase space variables, $n_1$ stands for 
the number of first class constraints, and $n_2$ stands for  number of second 
class constraints.\\

\indent
One can, however  infer the number of physical degrees of freedom for a system made up  
of  material medium, with  lesser effort, if one considers  taking the infrared limit we discussed
 earlier. In this limit the full in medium effective  Lagrangian ($L_{eff(m)}$) takes the form,

\begin{eqnarray}
L_{eff(m)}= -\frac{1}{4}F_{\mu\nu}F^{\mu\nu}+\frac{1}{2}\omega^{2}_{p}A_{0}A_{0},
\end{eqnarray}

\noindent
barring the Lorentz structure, this Lagrangian is very close to the Lagrangian of the Proca model. This model although is 
 gauge non-invariant, but known to produce correct number of degrees of freedom present
 in a massive theory.  This system  is known to have two second class constraints and no first class 
constraints. Therefore, the number of physical degrees of freedom, for the same,  turns out to be
three. Hence one needs to remove one of the four components of the  gauge potential, of U(1)
 gauge theory,  to  describe the dynamics of the system. We expect the same to hold good in our case.
The same is performed  in the next paragraph.  \\

\subsection{Gauge fixing:}

In continuation to the discussion presented in the last paragraph, we consider $A_{gf}$, that appears  in the equation for the gauge potential, 
\begin{eqnarray}
A^{\nu}(k)= A_{\parallel}(k) \hat{b}^{(1)\nu}+ A_{\perp}(k) \hat{I}^{\nu}+ A_{L}(k)
\hat{\tilde{u}}^{\nu} +  A_{gf}(k)\hat{k}^{\nu},
\label{gauge-pot4}
\end{eqnarray}

\noindent
 to be a redundant DOF and consider it to be equal to zero. As a consequence, it turns out that $k_{\nu}A^{\nu}(k)=0$ that  happens to be the Lorentz gauge condition. We would like to point out at this stage that basis vectors $b^{(1)\nu}$, $I^{\nu}$ and $\tilde{u}^{\nu}$  used in equation (\ref{gauge-pot2}) to describe the gauge potential $A^{\nu}(k)$ can always be rotated to a set of new basis vectors, however
the associated DOF in that basis may not be suitable for normal mode analysis of the system.

\subsection{ Some Interesting observations:}

We further note here that equations (\ref{ortho-vectors}) and (\ref{normalizators}) offer some interesting possibilities:
in terms of them  one can further define an effective metric in the momentum space as:
\begin{eqnarray}
G_{\mu\nu} = \frac{k_{\mu} k_{\nu}}{k^2}-  \frac{b^{(1)}_{\mu} b^{(1)}_{\nu}}{b^{(1)}_{\alpha} b^{(1)\alpha}} -  \frac{I_{\mu} I_{\nu}}{I_{\alpha}I^{\alpha}} -
 \frac{\tilde{u}_{\mu} \tilde{u}_{\nu}} {\tilde{u}_{\alpha} \tilde{u}^{\alpha}};  
\label{efft-metric}
\end{eqnarray}
using the orthogonal properties of the unit vectors it would be possible to 
to raise or lower the index of any general four-vector in momentum space using equation (\ref{efft-metric}). 
The other important consequence that follows from the definition of the vector potential given by equation (\ref{gauge-pot2}) is that, the gauge 
fixed  4-vector potential is space like  irrespective of the choice of the momentum.

\subsection{Discrete symmetries}

The discrete symmetry ({\bf C}, {\bf P} and {\bf T}) transformation  properties of the
tensors, four vectors and the EM form-factors, used here are  listed in 
table~[\ref{table:1}]. These transformation laws (agrees with ones provided in \cite{cherkas,jack,Itzykson-Zuber,Adler}) can be obtained
using standard QFT based arguments, except for $u^{\mu}$( which  can be obtained following the subtle principles of finite temperature field theory 
\cite{shahbad}).

\begin{table}[h!]
\centering
\begin{tabular}{||c | c c c c c c c c c c c c c ||} 
 \hline
         & $F_{\mu\nu}$     & $k_{\mu}$     &   $u_{\mu}$ &    $\tilde{u}_{\mu}$ & $b^{(1)}_{\mu}$ & $b^{(2)}_{\mu}$ & $ I_{\mu}$ & $A_{\parallel}$ & $ A_{\perp} $ &  $A_{L}$  &$f_{\mu\nu}$ & $i$ & $\epsilon_{\mu\nu\rho\sigma}$\\ [0.5ex] 
 \hline\hline
 {\bf C} & $-F_{\mu\nu}$    & $k_{\mu}$     & $-u_{\mu}$ &  $ -\tilde{u}_{\mu}$   & $-b^{(1)}_{\mu}$  & $-b^{(2)}_{\mu}$  &   $-I_{\mu}$  &  $A_{\parallel}$    & $  A_{\perp}  $ &  $A_{L}$ & $-f_{\mu\nu}$ & $i$ & $\epsilon_{\mu\nu\rho\sigma}$    \\ 
 {\bf P} & $ F^{\mu\nu}$    & $k^{\mu}$     &   $u^{\mu}$ &  $ ~\tilde{u}^{\mu}$   & $b^{(1)\mu}$ & $ b^{(2)\mu}$ &    $I^{\mu}$  & $ A_{\parallel}$    & $  A_{\perp}  $ &  $ A_{L}$   & $f_{\mu\nu}$ & $i$ & $-\epsilon_{\mu\nu\rho\sigma}$  \\
 {\bf T} & $-F^{\mu\nu}$    & $k^{\mu}$     &  $-u^{\mu}$ &    $-\tilde{u}^{\mu}$  & $ -b^{(1)\mu}$ & $ -b^{(2)\mu}$   &   $-I^{\mu}$  & $ - A_{\parallel} $ &  $ - A_{\perp}$ &  $ - A_{L} $ &$-f_{\mu\nu}$ & $-i$ & $-\epsilon_{\mu\nu\rho\sigma}$ \\ [1ex]
 
\hline
\hline
\end{tabular}
\caption{ Transformation properties for the vectors, tensors and the EM form-factors used to describe $A^{\nu}(k)$ in equation (\ref{gauge-pot2}), under ${\bf C}$, ${\bf P}$ and ${\bf T}$.}
\label{table:1}
\end{table}

\indent
The question that one would like to pose next is, how the interaction between 
charge neutral spin zero scalar with  charge neutral spin one 
photon, does take place. 
The best way to address the same is to take a discrete 
symmetry based route, done elegantly by  Raffelt and Stodolosky in \cite {Raffelt}, invoking {\bf CP} symmetry based arguments. Where  in the  case of 
a pseudo-scalar-photon (a$\gamma$) interacting system, it was shown that, 
in such situation, the {\bf CP} asymmetric pseudo-scalar (axion) would couple 
to the {\bf CP} asymmetric part of the photon's four vector potential. The 
{\bf CP } symmetric part  of the photon would remain decoupled.
That is the {\bf CP} violating helicity state (HS) of photon would
 couples with the
{\bf CP } violating axion field and  evolve in space and time; and the {\bf CP} preserving  helicity state of the photon propagates freely.  The  dilaton 
photon  dynamics, in similar situation, can be cast in a similar language as 
that of\cite {Raffelt},  exchanging the role of pseudo-scalars with scalars  
and scalars with pseudo-scalars.\\

\indent 
Instead of following \cite {Raffelt}, we describe  such a system here in terms 
of a set of EM  form-factors of the photon, described by equation (\ref{gauge-pot2}), introduced originally in \cite{Ganguly-jain}.
 The proof of these transformations properties are 
provided in the  supplementary document of this article. Having these 
transformation laws in hand, we argue  here, that, the equations of motion, 
when cast in terms of the electromagnetic form factors of 
table~[\ref{table:1}], follow a {\bf PT}  symmetry based coupling dynamics, 
instead of the {\bf CP} symmetry based one of \cite{Raffelt} . \\

\section{Incorporation of matter effects}

Matter effects are incorporated   through the inclusion of a term  of the following form, $A^{\mu}(k)\Pi_{\mu\nu}(T, \mu)A^{\nu}(k)$ in 
the effective  Lagrangian ($L_{eff}$) of the system  
\cite{Raffelt-magnetized,Pankaj-Ralston,Ganguly-ann,Ganguly-jain,Tercas,subir-axion,choi-kim, ganguly-konar-pal}.
This  is a scalar made up by contracting   in-medium photon self-energy 
tensor $\Pi_{\mu\nu}(T, \mu)$   with gauge fields. Parameters  T and $\mu$   
stand for temperature and chemical potential as arguments  of   
$\Pi_{\mu\nu}(T, \mu)$. The contribution of  the parity violating part of  weakly  magnetized matter effects is similarly taken into account by  the inclusion of a term like   $A^{\mu}(k)\Pi^{p}_{\mu\nu}(k,\mu,T,eB)A^{\nu}(k)$, when $\Pi^{p}_{\mu\nu}(k,\mu,T,eB)$ is the photon polarization tensor evaluated by incorporating the effects of the magnetic field to first order in the external  field 
strength $eB$ but exact   to all orders in
 T, and chemical potential $\mu$; using Schwinger's proper time propagator 
and formalism of finite temperature field theory \cite{ganguly-konar-pal,shahbad,nieves}. The photon polarization tensor $\Pi^{p}_{\mu\nu}(k)$ can be parameterised in the following way;

\begin{eqnarray}
\Pi^{p}_{\mu\nu} (k) = \Pi^{p}(k)P_{\mu\nu},
\hbox{ when~~~~~~~}
P_{\mu\nu} = i\epsilon_{\mu\nu\alpha{\beta}_{\parallel}}\frac{k^\alpha}{|k|}u^{\tilde{\beta}_{\parallel}},
\end{eqnarray}
\noindent
where $\epsilon_{\mu\nu\alpha{\beta}_{\parallel}}$ is Levi-Civita tensor and subscript $\beta_{\parallel}= 0,3$. In the expression for $P_{\mu\nu}$, $\tilde{\beta}_{\parallel}$ is defined such that if $\beta_{\parallel} = 0$ then $\tilde{\beta}_{\parallel} = 3$ and vice-versa. We have discussed the  {\bf C}, {\bf P} and {\bf T} symmetries  of polarization tensors in the Supplementary materials part of this article.  

\section{Equations of motion: }

Coupling between different DOF  of the system, follows from the form of the
effective Lagrangian.
 As can be verified from the equation  (\ref{L-total})  that, the couplings between  {\bf PT} violating  
$A_{\parallel}(k)$ and {\bf PT} symmetric $\phi(k)$, is generated in the 
$\phi\gamma$ tree level Lagrangian (in external $B$ field), through  a 
multiplicative {\bf T} violating factor $i$. Thus making the Lagrangian
{\bf PT} symmetric. The other two {\bf PT} violating EM form-factors $A_{\perp}(k)$ 
and $A_{L}(k)$  (see table~[\ref{table:1}]) have no coupling with the scalar $\phi(k)$. Therefore, with the inclusion of $\Pi_{\mu\nu}(T,\mu)$, when $A_{L}(k)$ becomes non-zero, the mixing matrix for $\phi\gamma$ remains $2\times2$. But with the inclusion of the, effective Lagrangian due to parity violating 
magnetized-media-induced photon self-energy term $\Pi^{p}(k)_{\mu\nu}$, 
$A_{\parallel}(k)$ gets coupled to $A_{\perp}(k)$, thus turning the 
mixing matrix  $3\times3$ one.\\

\indent
In magnetized media the effective  Lagrangian-- for $\phi\gamma$  
interactive system including the effective interaction Lagrangians  due to 
photon self-energy terms-- is given by:
\begin{eqnarray}
L_{eff,\phi} = \frac{1}{2}\phi[k^2-m_{\phi}^2]\phi -\frac{1}{4}f_{\mu\nu}f^{\mu\nu} + \frac{1}{2}A^{\mu}\Pi_{\mu\nu}A^{\nu} +\frac{1}{2} A^{\mu}\Pi^{p}_{\mu\nu}A^{\nu} 
-\frac{1}{4}g_{\phi\gamma\gamma}\phi{\bar{ F}}^{\mu\nu}f_{\mu\nu}.
\label{L-total}
\end{eqnarray}

\noindent
In eqn. (\ref{L-total}), the variable $f_{\mu\nu}$ stands for field strength for the dynamical photons,  $\bar{F}_{\mu\nu}$ stands for the external field, 
$\Pi_{\mu\nu}$ photon polarization tensor in an isotropic medium, and 
$\Pi^{p}_{\mu\nu}$ the same in presence of magnetic field to $O(eB)$.

\subsection{Mixing dynamics of $\phi\gamma$ interaction} 
The equations of motion for the EM form-factors for the photon,
in the notation  of \cite{Ganguly-jain}, turn out to be,\\

\begin{eqnarray}
 (k^2-\Pi_T)A_{\parallel} (k)+ i\Pi^{p}(k)N_{1}N_{2}\left[\epsilon_{\mu\nu\delta\beta}
\frac{k^{\beta}}{\mid{k}\mid}u^{\tilde{\delta}_{\parallel}} b^{(1)\mu} I^{\nu}\right] A_{\perp}(k)
&=& \frac{ig_{\phi\gamma\gamma}\phi(k)}{N_1},
\label{fd1} \\
 (k^2-\Pi_T)A_{\perp} (k) - i\Pi^{p}(k)N_{1}N_{2}\left[\epsilon_{\mu\nu\delta\beta}
\frac{k^{\beta}}{\mid{k}\mid}u^{\tilde{\delta}_{\parallel}} b^{(1)\mu} I^{\nu}\right]A_{\parallel}(k)&= &0,
\label{fd2} \\
(k^2 -\Pi_L)A_L(k) &=& 0. 
\label{fd3}
\end{eqnarray}   
\noindent
The three equations (i.e., (\ref{fd1})-(\ref{fd3})), describe the dynamics of 
the three DOF of the photon.   And the 
equation of motion for $\phi(k)$ is given by,
\begin{eqnarray}
(k^2 - m^2)\phi(k) = -\frac{ig_{\phi\gamma\gamma} A_{\parallel}(k)}{N_1}. 
 \label{fd4}
\end{eqnarray}

\noindent
It can be checked that, left and right sides of the  equations of motion 
above are {\bf PT}   symmetric. Once the discrete transformations {\bf P} and {\bf T} 
are applied on the variables on both side including
 background  field ($\bar{F}^{\mu\nu}$), EM form-factors of the  photons and  the scalar field $\phi(k)$, then it would show that equations of motion  (\ref {fd1} to \ref {fd4}) remain invariant.
For notational convenience, we next introduce the new variables $F$ and $G$, 
defined as,
\begin{eqnarray}
F =N_{1}\,N_{2}\, \Pi^{p}(k)\left[\epsilon_{\mu\nu\delta\beta}
\frac{k^{\beta}}{\mid{k}\mid}u^{\tilde{\delta}_{\parallel}} b^{(1)\mu} I^{\nu}\right] 
\mbox{~~and~~}  G = \frac{g_{\phi\gamma\gamma}}{N_1}.
\label{F-G}
\end{eqnarray}
\noindent
and would be using them when necessary.\\
\indent
 As stated already, in the long wavelength limit, we 
consider $\Pi_{T}=\omega^{2}_{p}$, where 
$\omega_p = \sqrt{\frac{4\pi\alpha n_e}{m_e}}$, is the plasma frequency, 
$\alpha$  is EM coupling constant and $n_e$ is the density of 
electrons. Other terms, $F$ and $G$ introduced in equation (\ref{F-G}) can be simplified 
to yield  $F=\frac{\omega^{2}_{p}}{\omega}\frac{eB\cos{{\theta}}}{m_e}$ and 
$G= - g_{\phi\gamma\gamma}B \sin\theta \omega$, where $\theta$ is angle between the 
photon propagation vector $\vec{k}$ and the magnetic field $B$.
 Here $m_e$ is the mass of 
electron and $e$ is the electronic charge.
The equations of motion can now be cast
in terms of a compact  $4 \times 4$  matrix  $ {\rm{\bf M}}^{\prime}$ as:\\  
\indent
\begin{eqnarray}
\left[\begin{array}{c} k^2 {\bf I} - {\bf M^{\prime} } 
 \end{array} \right]
\left(\begin{array}{c} A_{\parallel}(k) \\ A_{\perp}(k)\\ A_{L}(k) \\\phi(k) \end{array} \right)=0,
\label{photon-scalar-mixing-matixxMprime}
\end{eqnarray}

\indent
As we have already mentioned that for $\phi\gamma$ system in  
magnetized media, the longitudinal DOF of photon does not mix with rest others, 
therefore we will exclude this term from the mixing matrix, and upon doing so,  
the  $4 \times 4 $ mixing matrix ${\mathbf M^{\prime}}$,  reduces to a 
 $3\times 3$ mixing matrix denoted by ${\bf M}$. For the sake of brevity, using  shorthand notations the equations 
of motions now can be written as follows:
\begin{equation}
\left[\begin{array}{c} k^2 {\bf I} - {\bf M } 
 \end{array} \right]
\left(\begin{array}{c} A_{\parallel}(k) \\ A_{\perp}(k) \\\phi(k) \end{array} \right)=0,
\label{photon-scalar-mixing-matixx}
\end{equation} 

\noindent
where the $3\times3$ mixing matrix for scalar-photon interaction is given as; 

\begin{equation}
{\bf M} =
    \left( \begin{array}{ccc}              
 \omega^2_p   &  \;\;  iF           & \;\; -iG        \\
-iF           & \;\; \omega^{2}_{p} & \;\;    0       \\
iG            &  \;\;  0            & \;\;    m^2_{\phi}  
   \end{array} \right).
\label{mixing-mat1}   
\end{equation}

\noindent
In order to get the Stoke parameters ${\bf I}(\omega,z), {\bf Q}(\omega,z), 
{\bf U}(\omega,z)$ and ${\bf V}(\omega,z)$ for the EM radiation, we need to get 
the solutions of the field  equations; this can be achieved by 
diagonalizing {\bf M}.

\subsection{Diagonalizing the 3 $\times$ 3 mixing matrix}

Our objective here is to obtain the analytical expression for the 
unitary matrix {\bf U}, that would diagonalize the hermitian matrix 
{\bf M} . We express the elements of the same ({\bf U}), using  
analytic algebraic  expressions. Using this matrix we carry out the 
numerical operations, to  estimates of observables, maintaining  a 
numerical accuracy of the order $\sim 10^{-9}$ or more for the identities
those the various intermediate expressions of interest need to satisfy
during the numerical evaluation of the form factors.\\
\indent
In order to obtain the elements of the  matrix {\bf U}, we need to solve 
the characteristic equation, obtained from Det $({\bf M}- \lambda_{j}{\bf I}) = 0$, 
and find the eigen values (roots) i.e., $\lambda_{j}$ ($j=1, 2$ and $3$) of {\bf M}. Then  use the same to find the 
corresponding eigen vectors. Finally, using the eigen vectors, construct 
the unitary matrix  {\bf U} that would diagonalize {\bf M}. The characteristic 
equation for this $3 \times 3$  hermitian matrix {\bf M}, for obvious reasons  
turns out to be a cubic equation, having real roots. 
The cubic equation, that follows from the characteristic equation can be  written, in 
terms of parameters b, c and d as,  

\begin{eqnarray}
\lambda_{j}^3 + b \lambda_{j}^2 +c \lambda_{j} + d = 0,
\end{eqnarray}
where  the parameters b, c and d are functions of the elements of mixing matrix {\bf M},  
denoted by:
\begin{eqnarray}  
 b &=& -(2\omega^2_p + m^2_{\phi}) \\
 c &=& \omega^4_p + 2 \omega^2_p m^2_{\phi}- \left( \frac{e{ B}_{\parallel}}{m_e}\frac{\omega^2_p}{\omega}\right)^2 - (g_{\phi\gamma\gamma}{ B}_{\perp}\omega)^2 \\
 d &=& -\left[\omega^4_p m^2_{\phi} - \left( \frac{e{B}_{\parallel}}{m_e}\frac{\omega^2_p}{\omega}\right)^2 m^2_{\phi}  -  (g_{\phi\gamma\gamma}{ B}_{\perp}\omega)^2 \omega^2_p \right]. 
\end{eqnarray}  
Next we introduce the variables $P$ and $Q$, when $P= \left(\frac{3c-b^2}{9}\right)$ and 
$2Q = \left(\frac{2b^3}{27} -\frac{bc}{3} +d \right)$; in terms of them,
the roots turn out to be, \\
\begin{equation}
\begin{array}{cc}
     \lambda_{1} =&   {\bf R} \cos \alpha + \sqrt{3} {\bf R} \sin \alpha -{ b}/3, \\
     \lambda_{2} =&   {\bf R} \cos \alpha - \sqrt{3} {\bf R }\sin \alpha -{ b}/3, \\
     \lambda_{3} =& \hskip -1.5cm -2{\bf R} \cos \alpha  -{ b}/3, 
\end{array} 
\mbox{~~with~~} 
  \left\{  \begin{array}{c}
                 \alpha  =\frac{1}{3} \cos^{-1} \left( \frac{Q}{{\bf R}^3} \right )\\
                \,\,\,\,\, {\bf R} = \sqrt{\left(-P \  \right )} {\bf sgn} \left( Q \right). 
     \end{array}
  \right. 
\label{exact-roots}
\end{equation}
\noindent
It should be noted that, in principle {\bf R} can be equal to $+\sqrt{(-P)}$ or $-\sqrt{(-P)}$. However, the ratio $(\frac{Q}{{\bf R}^{3}})$ should be positive. Hence to maintain the same, the factor of ${\bf sgn }(Q)$ is introduced in the definition of {\bf R}. The  orthonormal eigenvectors  ${\bf X}_j$ of  {\bf M} are  to be found from the matrix relation,  $\left[ {\bf M}-\lambda_j{\bf I}\right] \left[ {\bf X}_{j}\right] = 0.$
 In terms of its elements, the normalized column vector  
$\left[ {\bf X}_{j}\right]$,  can be denoted as,
\begin{equation}
 \left[ {\bf X}_j\right] =  \left[\begin{array}{c}
                                         \bar{u}_{j} \nonumber \\
                                         \bar{v}_{j} \nonumber \\
                                         \bar{w}_{j} \nonumber \\
                                 \end{array}
                          \right].
\end{equation}

\indent
Following standard methods, one can evaluate these elements, in terms of the roots $\lambda_j$ and elements of the 
mixing matrix {\bf M}. The same, once evaluated turns out to be, 
\begin{equation}
\begin{array}{cc}
     \bar{u}_{j} =&  ({\omega^2_p} -\lambda_j)(m_{\phi}^2 -\lambda_j) \times  \rm{\cal{N}}^{(j)}_{vn}, \\
    \bar{v}_{j} =&  i\frac{e{ B}_{\parallel}}{m_e}\frac{\omega^2_p}{\omega}(m_{\phi}^2 -\lambda_j)  \times  \rm{\cal{N}}^{(j)}_{vn}, \\
     \bar{w}_{j} =&  ig_{\phi\gamma\gamma}{ B}_{\perp}\omega({\omega^2_p} -\lambda_j)  \times  \rm{\cal{N}}^{(j)}_{vn} , 
\end{array} 
{\mbox{~~when~~}} 
  \left\{  \begin{array}{c}
                 \rm{\cal{N}}^{(j)}_{vn}  = \frac{1}{\sqrt{[|\bar{u}_j|^2 + |\bar{v}_j|^2 +  |\bar{w}_j|^2]}}.  \\
                   \end{array}
  \right.
\label{exact-eigen-vectors-3x3}
\end{equation}

Here $\rm{\cal{N}}^{(j)}_{vn}$  is normalisation constant and $j$ can take values from $1$ to $3$. Using these eigenvectors, the unitary matrix {\bf U} turns out to be :

\begin{equation}
{\bf U} =  
               \left( \begin{array}{ccc} 
(\omega^2_p -\lambda_1)(m^2_{\phi} - \lambda_1) \rm{{\cal{N}}}^{(1)}_{vn}                                   &  \hskip .5cm    (\omega^2_p -\lambda_2)(m^2_{\phi} - \lambda_2)  \rm{{\cal{N}}}^{(2)}_{vn}                                      &  \hskip .5cm   (\omega^2_p -\lambda_3)(m^2_{\phi} - \lambda_3)\rm{{\cal{N}}}^{(3)}_{vn}  \\
i\frac{e{B}_{\parallel}}{m_e}\frac{\omega^2_p}{\omega} (m^2_{\phi} - \lambda_1)\rm{{\cal{N}}}^{(1)}_{vn}    &  \hskip .5cm    i\frac{e{B}_{\parallel}}{m_e}\frac{\omega^2_p}{\omega} (m^2_{\phi} - \lambda_2) \rm{{\cal{N}}}^{(2)}_{vn}       & \hskip .5cm  i\frac{e{ B}_{\parallel}}{m_e}\frac{\omega^2_p}{\omega} (m^2_{\phi} - \lambda_3)\rm{{\cal{N}}}^{(3)}_{vn} \\
 ig_{\phi\gamma\gamma}{B}_{\perp}\omega(\omega^2_p -\lambda_1)\rm{\cal{N}}^{(1)}_{vn}                     & \hskip .5cm  ig_{\phi\gamma\gamma}{B}_{\perp}\omega (\omega^2_p -\lambda_2) \rm{\cal{N}}^{(2)}_{vn}                           & \hskip .5cm  ig_{\phi\gamma\gamma}{B}_{\perp}\omega (\omega^2_p -\lambda_3)\rm{\cal{N}}^{(3)}_{vn}
           \end{array} 
  \right). \label{u-mat}
\end{equation} 
\indent
The  unitary matrix given by (\ref{u-mat}) is the one that diagonalizes the 
3$\times$3 mixing matrix $\bf{M}$.
\subsection{Field equation : Solutions}
 In order to obtain the solutions of the coupled equation (\ref{photon-scalar-mixing-matixx}), one can   multiply the same by inverse of the matrix given by eqn.(\ref{u-mat})
 i.e., {\bf{U}$^{-1}$} from left and use the property of unitary matrices ${\bf UU}^{-1}= {\bf1}$ in the same equation 
(i.e., (\ref{photon-scalar-mixing-matixx})), and arrive at,  

\indent
\begin{equation}
{\bf{U}}^{-1}\left[\begin{array}{c} k^2 {\bf I} - {\bf M } 
 \end{array} \right]{\bf{UU}}^{-1}
\left(\begin{array}{c} A_{\parallel}(k) \\ A_{\perp}(k) \\\phi(k) \end{array} \right)=
\left[\begin{array}{c} k^2 {\bf I} - {\bf M}_{D} 
 \end{array} \right]
\left(\begin{array}{c} A'_{\parallel}(k) \\ A'_{\perp}(k) \\\phi'(k) \end{array} \right)=0.
\label{mat-eq-a}
\end{equation} 

\noindent
In order to arrive at  equation (\ref {mat-eq-a}), we have used the following notation,
\begin{eqnarray}
 \left(\begin{array}{c} A'_{\parallel}(k) \\ A'_{\perp}(k) \\ \phi'(k) 
\end{array} \right)={\bf U}^{-1}
\left(\begin{array}{c} A_{\parallel}(k) \\ A_{\perp}(k) \\\phi(k) \end{array} \right).
\label{primed-degrees-of-freedom}
\end{eqnarray}

\noindent
The Matrix ${\bf M}_D$ is the diagonal matrix given by ${\bf M}_{D}={\bf U}^{-1}{\bf M} {\bf U} $  that has  eigen values of the matrix {\bf M} as diagonal elements.
For a propagating beam of photons in the $z$ direction, $k_{3}$ can be 
re-transformed back to $z$ by  taking inverse Fourier transform. Furthermore 
one can  express $k^2 \approx 2\omega(\omega -i\partial_z)$, without much 
loss of generality, to use in the equations of motion.  As a result of 
these (algebraic manipulations, equation (\ref{mat-eq-a}) assumes the 
following form,  

\begin{equation}
\left[\begin{array}{c}  (\omega -i\partial_z)  \mathbf{I}  - \left[ \begin{array}{ccc} \frac{\lambda_1}{2\omega} &   0     &    0 \\
0 &  \frac{\lambda_2}{2\omega}  & 0 \\
0 & 0 & \frac{\lambda_3}{2\omega}  \end{array} \right]        \end{array} \right]
\left[\begin{array}{c} {A'}_{\parallel}(z)\\ {A'}_{\perp}(z) \\ {\phi'}(z) \end{array} \right]=0.
\label{dmat}
\end{equation} 
\noindent
It is now easy to solve the matrix equation  (\ref{dmat}) by introducing 
 the variables, 
$\Omega_\parallel = \left(\omega - \frac{\lambda_1}{2\omega}\right) $,
$\Omega_\perp = \left(\omega - \frac{\lambda_2}{2\omega}\right)$ 
and $\Omega_\phi = \left(\omega - \frac{\lambda_3}{2\omega}\right)$. Instead of going into details,
we can now directly write down the solutions for the column vector 
$\left[{\mathbf A}(z), \phi(z) \right]^{T}$; in  matrix form, they are given by:
%
\begin{equation}
  \left[
\begin{matrix}
{{A}_{\parallel}(z)} \cr
{{A}_{\perp}(z)} \cr   
{\phi(z)}
\end{matrix}
\right] = {\bf U}\left[ \begin{array}{ccc} e^{-i\Omega_\parallel z} &   0    &    0 \\
0 & e^{-i\Omega_\perp z}  & 0 \\
0 & 0 & e^{-i\Omega_\phi z}  \end{array} \right] {\bf U}^{-1} 
\left[
\begin{matrix}
{{A}_{\parallel}(0)} \cr
{{A}_{\perp}(0)} \cr   
{\phi(0)}
\end{matrix}
\right]
\label{solnmat}.
\end{equation}
\noindent
The magnitudes of the elements of column vector 
$\left[{\mathbf A}(0), \phi(0) \right]^{T}$ in 
equation (\ref{solnmat}), are subject to the boundary conditions appropriate
for the physical situations assumed to be prevailing at the origin. Using 
the same initial conditions,  one can write down the 
solution of equation (\ref{solnmat}) for {A$_\parallel$}($\omega, z$), 
and it is:
\begin{eqnarray}
{ A_{\parallel}(\omega,z)}&=& 
\left( e^{-i\Omega_\parallel z }{\bar u}_1 { \bar u}^{*}_1  +  e^{-i\Omega_\perp z } {\bar u}_2{ \bar u}^{*}_2   
+ e^{-i\Omega_\phi z } {\bar u}_3{ \bar u}^{*}_3 \right){A_{\parallel}(\omega, 0)}  \nonumber \\
&+& \left(
e^{-i\Omega_\parallel z }{\bar u}_1 { \bar v}^{*}_1  +  e^{-i\Omega_\perp z } {\bar u}_2{ \bar v}^{*}_2   
+ e^{-i\Omega_\phi z } {\bar u}_3{ \bar v}^{*}_3 \right){ A_{\perp}(\omega, 0)}. 
\label{soln-a-parallel}
\end{eqnarray}  
\noindent 
Similarly the perpendicular component $A_{\perp}(\omega,z) $, turns out to be, 
\begin{eqnarray}
{ A_{\perp}(\omega,z)}&=& 
\left( e^{-i\Omega_\parallel z }{\bar v}_1 { \bar u}^{*}_1  +  e^{-i\Omega_\perp z } {\bar v}_2{ \bar u}^{*}_2   
+ e^{-i\Omega_\phi z } {\bar v}_3{ \bar u}^{*}_3 \right){ A_{\parallel}(\omega, 0)}  \nonumber \\
&+& \left(
e^{-i\Omega_\parallel z }{\bar v}_1 { \bar v}^{*}_1  +  e^{-i\Omega_\perp z } {\bar v}_2{ \bar v}^{*}_2   
+ e^{-i\Omega_\phi z } {\bar v}_3{ \bar v}^{*}_3 \right){ A_{\perp}(\omega, 0)}.  
\label{soln-a-perp}
\end{eqnarray}  

\noindent
Here, the parameters ${ A_{\perp}(\omega, 0)}$ and ${ A_{\parallel}(\omega, 0)}$ present in equation (\ref{soln-a-parallel}) and  (\ref{soln-a-perp}) are  ${ A_{\perp}(\omega, z)}$ and ${ A_{\parallel}(\omega, z)}$ respectively under the boundary conditions as mentioned above. 

\section{Polarimetric observables }
From the coherency matrix, Stokes parameters, can be obtained. 
In terms of the solutions of the field equations they can be expressed as: 
 \begin{eqnarray}
\!\!{\bf I}(\omega, z)&=& <{A_{\parallel}}^{*}(\omega, z)A_{\parallel}(\omega, z)>+<A_{\perp}^{*}(\omega, z)A_{\perp}(\omega, z)>, 
\nonumber  \\
{\bf Q}(\omega, z)&=&  <{A_{\parallel}}^{*}(\omega, z)A_{\parallel}(\omega, z)>-<A_{\perp}^{*}(\omega, z)A_{\perp}(\omega, z)> ,
\nonumber \\
{\bf U}(\omega, z)&=&2 Re <{A_{\parallel}}^{*}(\omega, z)A_{\perp}(\omega, z)>,
\nonumber \\
{\bf V}(\omega, z)&=& 2Im <{A_{\parallel}}^{*}(\omega, z)A_{\perp}(\omega, z)>.
\label{stokes-formal} 
\end{eqnarray}
\indent
It should be noted that ${\bf V}(\omega, z)$ in the equation (\ref{stokes-formal}) 
is a measure of circular polarization.  Other polarimetric observables, 
i.e., degree of linear polarization, ellipticity angle, polarization angle,  
follows from  the expressions of  ${\bf I}(\omega, z)$, ${\bf Q}(\omega, z)$, 
${\bf U}(\omega, z)$ and ${\bf V}(\omega, z)$.
The degree of linear polarization (represented as $P_{L}$) along with $\Pi$, 
are given by, 
\begin{eqnarray}
P_{L} &=& \frac{\sqrt{{\bf Q}^2(\omega,z) + {\bf U}^2(\omega,z)}}{{\bf I}(\omega,z)},\\
\label{pl} 
\Pi &=& \frac{ {\bf Q}(\omega,z)}{{\bf I}(\omega,z)}
\label{Pi}
\end{eqnarray}

\noindent
The polarization angle  (represented by $\Psi_p$),   is defined  in terms of 
${\bf U}$ and ${\bf Q}$, as:, 
\begin{eqnarray}
\tan(2\Psi_p) = \frac{{\bf U}(\omega,z)}{{\bf Q}(\omega,z)}.
\label {pol_angle}
\end{eqnarray}
 The ellipticity angle (denoted by $\chi$) , is defined as:
\begin{eqnarray}
\tan(2{ \chi}) = \frac{{\bf V}(\omega,z)}{\sqrt{{\bf Q}^2(\omega,z) + {\bf U}^2(\omega,z)}}.
\label{chi}
\end{eqnarray}

\subsection{Stokes ${\bf I}(\omega, z)$ and ${\bf Q}(\omega, z)$}
\indent
 To find out the expressions of Stokes parameters {\bf I}($\omega$, z) and {\bf Q}($\omega$, z)   for Scalar-photon mixing , we need to evaluate  ${{|A_{\parallel}(\omega,z)|^2}}$ and ${{|A_{\perp}(\omega,z)|^2}}$  , 
using equations (\ref{soln-a-parallel}) and (\ref{soln-a-perp}). 
Introducing the new variables, $\mathbb{P} = |{\bar u}_1||{\bar v}_1|$, 
$\mathbb{Q} = |{\bar u}_2||{\bar v}_2|$ and $\mathbb{R} = |{\bar u}_3||{\bar v}_3|$, the expression for 
${{|A_{\parallel}(\omega,z)|^2}}$ in terms of them, turns out to be,

\begin{align}
{{|A_{\parallel}(\omega,z)|^2}} =  \Big[1 
- &4|{\bar{u}_1}|^2|{\bar u}_2|^2\; \sin^2 \left(\frac{(\Omega_{\parallel}-\Omega_{\perp})z}{2} \right)
- 4|{\bar u}_2|^2|{\bar u}_3|^2\; \sin^2 \left(\frac{(\Omega_{\perp} -\Omega_{\phi})z}{2} \right)\nonumber\\  
- &4|{\bar u}_3|^2|{\bar u}_1|^2\; \sin^2 \left(\frac{(\Omega_{\phi} -\Omega_{\parallel})z}{2} \right)\Big]
\times { |A_{\parallel}(\omega,0)|^2}\nonumber\\ 
- &4 \Big[\mathbb{PQ}\; \sin^2 \left(\frac{(\Omega_{\parallel} -\Omega_{\perp})z}{2} \right) 
+ \mathbb{QR}\; \sin^2 \left(\frac{(\Omega_{\perp} -\Omega_{\phi})z}{2} \right)\nonumber\\  
+ &\mathbb{RP}\; \sin^2 \left(\frac{(\Omega_{\phi} -\Omega_{\parallel})z}{2} \right)\Big]
\times  { |A_{\perp}(\omega,0)|^2 }\nonumber\\
+ \Big[&|{\bar u}_1 {\bar u}_2|(|{\bar u}_1{\bar v}_2|- |{\bar u}_2{\bar v}_1|)\sin \left({(\Omega_{\parallel} -\Omega_{\perp})z} \right) \nonumber\\
+& |{\bar u}_2 {\bar u}_3|(|{\bar u}_2{\bar v}_3|-|{\bar u}_3{\bar v}_2|)\sin \left({(\Omega_{\perp} -\Omega_{\phi})z} \right) \nonumber\\
+&|{\bar u}_3 {\bar u}_1|(|{\bar u}_3{\bar v}_1|-|{\bar u}_1{\bar v}_3|)\sin \left({(\Omega_{\phi} -\Omega_{\parallel})z} \right)\Big]\nonumber\\
\times &2{|A_{\parallel}(\omega,0)|}{|A_{\perp}(\omega,0)|}.
\label{A_parallel-square}
\end{align}


\noindent 
Similarly, we can find ${{|A_{\perp}(\omega,z)|^2}}$. The expression for the same, in terms of the variables introduced earlier, turns out to be, 

\begin{align}
{{|A_{\perp}(\omega,z)|^2}} =- 4\Big[&\mathbb{PQ}\sin^2 \left(\frac{(\Omega_{\parallel} -\Omega_{\perp})z}{2} \right) + \mathbb{QR}\; \sin^2 \left(\frac{(\Omega_{\perp} -\Omega_{\phi})z}{2} \right)\nonumber\\ 
+&\mathbb{RP}\; \sin^2 \left(\frac{(\Omega_{\phi} -\Omega_{\parallel})z}{2} \right)\Big]
\times { |A_{\parallel}(\omega,0)|^2} \nonumber\\
+ \Big[1 -& 4|{\bar v}_1|^2|{\bar v}_2|^2 \;\sin^2 \left(\frac{(\Omega_{\parallel} 
-\Omega_{\perp})z}{2} \right)
- 4|{\bar v}_2|^2 |{\bar v}_3|^2\; \sin^2 \left(\frac{(\Omega_{\perp} -\Omega_{\phi})z}{2} \right)\nonumber\\  
- &4|{\bar v}_3|^2|{\bar v}_1|^2\; \sin^2 \left(\frac{(\Omega_{\phi} -\Omega_{\parallel})z}{2} \right)\Big]
\times { |A_{\perp}(\omega,0)|^2}\nonumber\\
+ \Big[&|{\bar v}_1 {\bar v}_2|(|{\bar u}_1{\bar v}_2|-|{\bar u}_2{\bar v}_1|)\sin \left({(\Omega_{\parallel} -\Omega_{\perp})z} \right)\nonumber\\
+& |{\bar v}_2 {\bar v}_3|(|{\bar u}_2{\bar v}_3|-|{\bar u}_3{\bar v}_2|)\sin \left({(\Omega_{\perp} -\Omega_{\phi})z} \right)\nonumber\\
+&|{\bar v}_3 {\bar v}_1|(|{\bar u}_3{\bar v}_1|-|{\bar u}_1{\bar v}_3|)\sin \left({(\Omega_{\phi} -\Omega_{\parallel})z} \right)\Big]\nonumber\\
\times & 2{ |A_{\parallel}(\omega,0)|}{|A_{\perp}(\omega,0)|}.
\label{A_perp-square}
\end{align}

\noindent
Using equations (\ref{A_parallel-square}) and (\ref{A_perp-square}), one can get the expressions for ${\bf I}(\omega, z)$ and ${\bf Q}(\omega, z)$ as follows:

\begin{eqnarray}
{\bf I}(\omega, z)& =& |A_{\parallel}(\omega, z)|^{2} +   |A_{\perp}(\omega, z)|^{2}\\
{\bf Q}(\omega, z)& = &|A_{\parallel}(\omega, z)|^{2} -   |A_{\perp}(\omega, z)|^{2}.
\label{IandQ}
\end{eqnarray}

\subsection{Stokes ${\bf U}(\omega, z)$ and ${\bf V}(\omega, z)$}
The expressions for ${\bf U}(\omega, z)$ and ${\bf V}(\omega, z)$ can be written in terms of $\cal{U}_{\parallel}$, $\cal{U}_{\perp}$, $\cal{U}_{\parallel \perp}$ and $\cal{V}_{\parallel}$, $\cal{V}_{\perp}$, $\cal{V}_{\parallel \perp}$. For ${\bf U}(\omega, z)$ they are given by :

\begin{align}
{\cal {U}_{\parallel}} &= |{\bar u}_1{\bar u}_2|(|{\bar v}_1{\bar u}_2| -|{\bar u}_1{\bar v}_2|)
\sin \left({(\Omega_{\parallel} -\Omega_{\perp})z} \right) 
+|{\bar u}_2{\bar u}_3|(|{\bar v}_2{\bar u}_3| -|{\bar u}_2{\bar v}_3|)
\sin \left({(\Omega_{\perp} -\Omega_{\phi})z}\right)  \nonumber\\
&+|{\bar u}_3{\bar u}_1|(|{\bar v}_3{\bar u}_1| -|{\bar u}_3{\bar v}_1|)
\sin \left({(\Omega_{\phi} -\Omega_{\parallel})z} \right). 
\end{align}
\begin{align}
{\cal {U}_{\perp}}&= |{\bar v}_1{\bar v}_2|(|{\bar v}_1{\bar u}_2| - |{\bar u}_1{\bar v}_2|)
\sin \left({(\Omega_{\parallel} -\Omega_{\perp})z} \right)
+|{\bar v}_2{\bar v}_3|(|{\bar v}_2{\bar u}_3| -|{\bar u}_2{\bar v}_3|)
\sin \left({(\Omega_{\perp} -\Omega_{\phi})z} \right) \nonumber\\
&+ |{\bar v}_3{\bar v}_1|(|{\bar v}_3{\bar u}_1| -|{\bar u}_3{\bar v}_1|)
\sin \left({(\Omega_{\phi} -\Omega_{\parallel})z} \right).
\end{align}
\begin{align}
\hskip -4cm{\cal {U}_{\parallel \perp}}& =  (|{\bar v}_1{\bar u}_2| - |{\bar u}_1{\bar v}_2|)^2 \;\cos \left({(\Omega_{\parallel} -\Omega_{\perp})z} \right)
+(|{\bar v}_2{\bar u}_3| - |{\bar u}_2{\bar v}_3|)^2\;\cos \left({(\Omega_{\perp} -\Omega_{\phi})z} \right)\nonumber\\ 
&+ (|{\bar v}_3{\bar u}_1| -|{\bar u}_3{\bar v}_1|)^2\;\cos \left({(\Omega_{\phi} -\Omega_{\parallel})z} \right). 
\end{align}
 And similarly, the expressions for ${\cal {V}_{\parallel }}$, ${\cal {V}_{\perp }} $ and ${\cal {V}_{\parallel \perp }}$, introduced to express the measure of circular polarization ${\bf V}(\omega, z)$  are:
\begin{align}
{\cal {V}_{\parallel}}&=  [(|{\bar u}_1 {\bar v}_1||{\bar u}_1|^2 + |{\bar u}_2 {\bar v}_2||{\bar u}_2|^2 + |{\bar u}_3{\bar v}_3||{\bar u}_3|^2) 
+ |{\bar u}_1{\bar u}_2|({|\bar v}_1{\bar u}_2| + |{\bar u}_1{\bar v}_2|)\;\cos \left({(\Omega_{\parallel} -\Omega_{\perp})z} \right)\nonumber\\ 
&+ |{\bar u}_2{\bar u}_3|(|{\bar v}_2{\bar u}_3| + |{\bar u}_2{\bar v}_3|)\;\cos \left({(\Omega_{\perp} -\Omega_{\phi})z} \right)
+|{\bar u}_3 {\bar u}_1|(|{\bar v}_3{\bar u}_1| +|{\bar u}_3{\bar v}_1|)\;\cos \left({(\Omega_{\phi} -\Omega_{\parallel})z} \right)].
\end{align}
\begin{align}
{\cal {V}_{\perp}}&= [(|{\bar u}_1 {\bar v}_1||{\bar v}_1|^2 + |{\bar u}_2 {\bar v}_2||{\bar v}_2|^2 + |{\bar u}_3 {\bar v}_3||{\bar v}_3|^2) 
+{|\bar v}_1{\bar v}_2|({|\bar u}_1{\bar v}_2| + |{\bar v}_1{\bar u}_2|)\;\cos \left({(\Omega_{\parallel} -\Omega_{\perp})z} \right)\nonumber\\ 
&+ {\bar v}_2{\bar v}_3|(|{\bar u}_2{\bar v}_3| + |{\bar u}_3{\bar v}_2|)\;\cos \left({(\Omega_{\perp} -\Omega_{\phi})z} \right)
+|{\bar v}_3 {\bar v}_1|(|{\bar u}_3{\bar v}_1| + |{\bar u}_1{\bar v}_3|)\;\cos \left({(\Omega_{\phi} -\Omega_{\parallel})z} \right)]. 
\end{align}
\begin{align}
\hskip-.7cm{\cal {V}_{\parallel \perp}}& = [(|{\bar u}_1|^2|{\bar v}_2|^2 - |{\bar u}_2|^2|{\bar v}_1|^2)\sin \left({(\Omega_{\parallel} -\Omega_{\perp})z} \right)
+(|{\bar u}_2|^2|{\bar v}_3|^2 - |{\bar u}_3|^2|{\bar v}_2|^2)
\sin \left({(\Omega_{\perp} -\Omega_{\phi})z} \right)\nonumber\\
&+(|{\bar u}_3|^2|{\bar v}_1|^2 - |{\bar u}_1|^2|{\bar v}_3|^2)
\sin \left({(\Omega_{\phi} -\Omega_{\parallel})z} \right)].
\end{align}

\noindent
Using them, one can write the expressions for  {\bf U}($\omega$, z) and {\bf V}($\omega$, z) as follows:

\begin{eqnarray}
{\bf U}(\omega, z) &=& {\cal{U}}_{\parallel} \times 2|A_{\parallel}(\omega , 0)|^{2} +{\cal{U}}_{\perp} \times 2|A_{\perp}(\omega , 0)|^{2}+{\cal{U}}_{\parallel \perp} \times 2|A_{\perp}(\omega , 0)||A_{\parallel}(\omega , 0)|\\
{\bf V}(\omega, z) &=& {\cal{V}}_{\parallel} \times 2|A_{\parallel}(\omega , 0)|^{2} +{\cal{V}}_{\perp} \times 2|A_{\perp}(\omega , 0)|^{2}+{\cal{V}}_{\parallel \perp} \times 2|A_{\perp}(\omega , 0)||A_{\parallel}(\omega , 0)|.
\label{calUV}
\end{eqnarray}

\noindent 
 The measures of linear and circular polarization {\bf Q}($\omega$, z), {\bf U}($\omega$, z) and polarization angle $\Psi_{p}$  are plotted as a function of energy  in figures  [\ref{f1}]. The discussion about them is provided below.

\section{ Possible Physical Situation}
Gamma ray bursts (GRB) are stellar sized explosions 
( those usually found with an associated luminosity  of about $\sim 10^{51}$ 
ergs/sec),  observed to occur isotropically at a red-shifts 
$z=1$ or $z=2$with the EM radiation in the form of X-ray or 
gamma-ray, with fluence of the order of $10^{-6}$  ergs/cm$^2$, as  observed
from earth \cite{packzynski}. Their size is usually estimated from
the time scale variability of their light curve,  estimated
to be around 0.1 sec \cite{ellis}.\\
\indent
 Accordingly their size is estimated 
to be  $\sim 10^9$cm. The energy is believed to be injected from an 
object, like neutron star, of radius $10^{6}$ cm. The high the degree of 
linear  polarization associated with the EM spectrum ( e.g. $ \Pi =  27\pm 11~ \%$ in GRBICO826A ), associated with them, makes one infer, 
a strong magnetic field  $\sim 10^{9}$ Gauss to be associated with them. 
The plasma frequency $\omega_p$ associated with them are believed  to lie
above, $\omega_p > 10^{-17} $GeV$ $ \cite {yonetokuI,yonetokuII,
rubbia}. The mechanism behind this explosion is currently under scrutiny; one believes 
that the polarization studies  with intensity of the spectra holds the
key to understand the geometry of the magnetic field at the source and the 
energy release mechanism. Though these aspects can be studied using classical 
physics, however one needs to be cautious, since factors like, presence of 
ALPs also potentially contribute to the modification of these observables. 
Therefore any anomaly in these signature may provide a clue to the existence of ALPs.

\section{Results}
\begin{figure}[h!]
\begin{minipage}[b]{.50\textwidth}  
\includegraphics[width=1\linewidth]{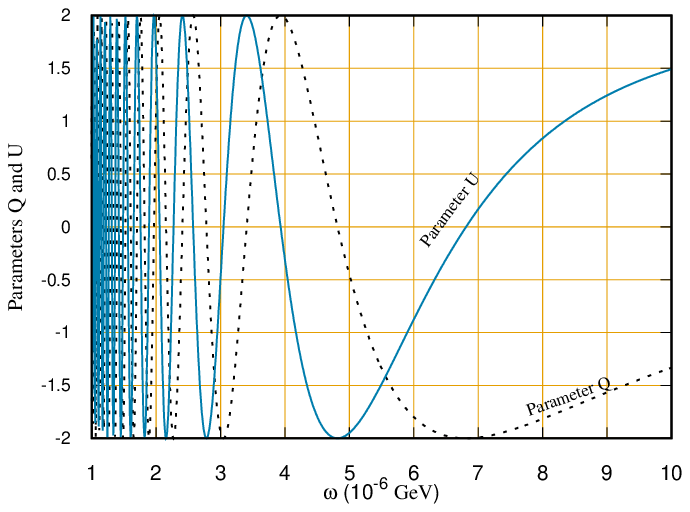}
\end{minipage}
\begin{minipage}[b]{.50\textwidth}  
\includegraphics[width=1\linewidth]{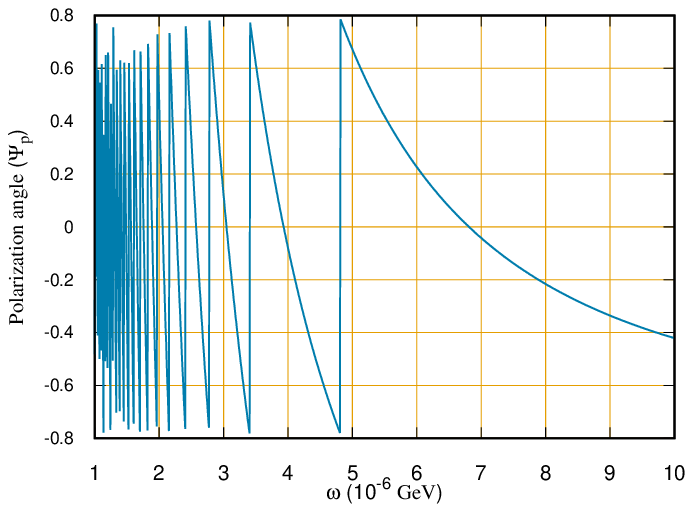}
\end{minipage}
\caption{  (In left), plot of Stokes parameter {\bf Q} and {\bf U} vs energy ($\omega$). (In right), plot of  polarization angle ($\Psi_{p}$)  vs energy ($\omega$). Here, $m_{\phi}$ = 1.0 $\times$ $10^{-12}$ GeV and $\omega_p$ = 3.7 $\times$$10^{-13}$GeV).}
\label{f1}
\end{figure}

Assuming a global magnetic field  of magnitude
$\sim 10^{9}$ Gauss, to be existing in a GRB fireball--
over a  path length of $\sim10^9$cm, the 
stokes parameters {\bf Q}, {\bf U} and  the polarization angle 
${\Psi_{p}}$ were estimated  using equation (\ref{stokes-formal})  
and (\ref {pol_angle}) numerically -- for scalar-photon coupling 
constant  $g_{\phi\gamma\gamma}\sim10^{-11}$GeV$^{-1}$  with 
$\omega _{p}\sim10^{-13}$GeV, and $m_{\phi}\sim10^{-2}$eV. 
The plots of the same are provided in figure [\ref {f1}].

\begin{figure}[ht!]
\begin{minipage}[b]{.50\textwidth}  
\includegraphics[width=1\linewidth]{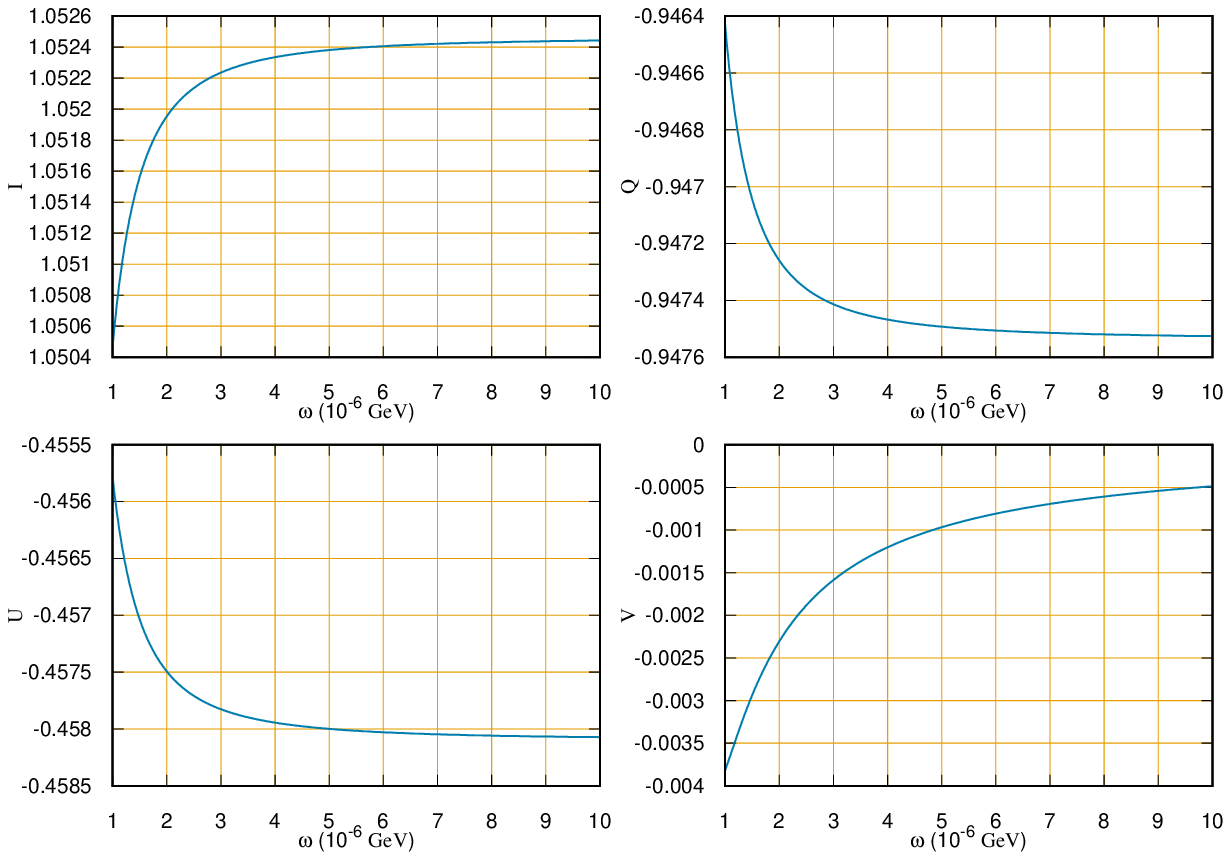}

 \end{minipage}
\begin{minipage}[b]{.50\textwidth}  
\includegraphics[width=1\linewidth]{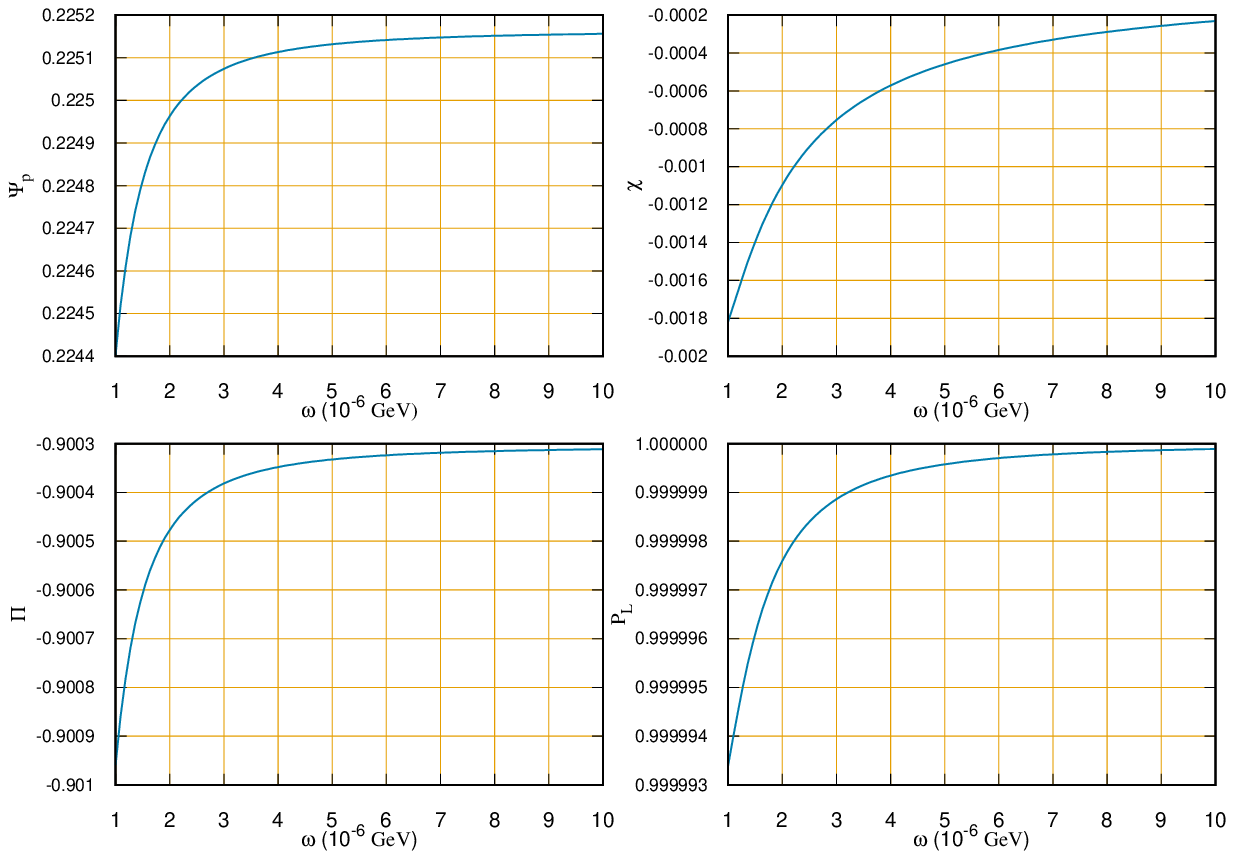}
\end{minipage}
\caption{ (In left panel) plots of Stokes parameters ({\bf I, Q, U , V}) vs energy ($\omega$). (In right panel) plots of polarization angle ( $\Psi_{p}$, ellipticity angle ($\chi$), degree of polarization ($\Pi$) and linear polarization ($P_{L}$) vs energy ($\omega$).  Here, $m_{\phi}$ = 1.0 $\times$ $10^{-15}$ GeV and $\omega_p$ = 3.7 $\times $ $10^{-15}$ GeV.}
\label{f2}
\end{figure}

\noindent
Since the degree of linear polarization $\Pi$, turns out to be equal to  {\bf Q}, 
when  {\bf U} is neglected (as is evident from the respective definitions of the same),  
we can get the information about the amplitude variation for {\bf Q}, { $\mathbf\Pi$} and ${\mathbf \Psi_{p}}$ 
over the energy ($\omega$) interval $1.0 \times 10^{-6} $GeV$< 
\omega < 1.0 \times 10^{-5}$GeV$ $ from the plots provided in figure [\ref{f1}].\\ 
\indent
In most of the astrophysically viable, satellite based experiments, 
the detectors are hardly line sensitive, they usually operate over a broad energy 
range; hence the parameters (of Stokes) are usually estimated by adding 
the signal strengths over an energy range (according to the detector 
under consideration), followed by an averaging of the signal -- over 
that same energy range.\\
\indent
Now if we look into these plots, we notice the existence of the highly
oscillating part for the estimates of {\bf Q} 
in the energy interval, $1.0 \times 10^{-6} $GeV$ <\omega<3.0 \times
10^{-6}$GeV$ $  and a similar pattern also for {\bf U} in the interval
 $1.0 \times 10^{-6}$GeV$<\omega< 2.0 \times 10^{-6}$GeV. This is followed by 
an monotonically increasing or decreasing pattern (left panel of 
figure [\ref{f1}]). Similar effect, is identifiable in the plot for 
{\bf $\Psi _{p}$} vs energy in the right panel of the same figure 
(i.e., figure [\ref{f1}]).\\
\noindent
Therefore if the standard satellite based experiment dictated data extraction 
prescriptions are followed -- while extracting signals from the \
datum like that  
producing figure [\ref {f1}]-it would lead to the  generation of an extremely 
low strength unphysical signal. This happens due to a major cancellation of 
the contributions due to strong oscillations of the actual signals around 
zero.\\ 
\indent
Thus, in order to get a realistic and statistically significant estimate of a
signal, from detectors operating over a broad energy range, one must explore the
available parameter space, over which the variations of the signals are stable
with energy.\\
\noindent
A collection of data sets for the spectro-polarimetric observables, those (unlike 
the ones of  figure [\ref {f1}]) looked stable  under the variation 
with respect  to $\omega)$, were found  in the 1-10 KeV range, when  
the numerical size  of the parameters $\omega _p$, and $m_{\phi}$ were 
considered at around $\sim10^{-15}$GeV, while the path length($z$), 
magnetic field strength($eB$) and scalar photon coupling strength
($g_{\phi\gamma\gamma}$)  were maintained at  $10^6$cm,  $eB  \sim 10^{9}$ Gauss 
and, $g_{\phi\gamma\gamma}\sim10^{-11}  $GeV$^{-1}$, respectively.\\
\indent
The plots of the spectro-polarimetric observables, such as  linear 
polarization ($P_L$), polarization angle $\Psi_p$, ellipticity angle $\chi$ 
and degree of polarization $\Pi$, estimated with these parameters, 
are displayed in figure [\ref {f2}], for an energy interval-- 
in the 1-10 KeV range.
Interestingly enough, the estimates of maximum linear polarization, 
for these parameters and energy range, is about $99\%$ 
and the polarization angle is about $12^{o}$; those lie well within 
the range of the observed linear-polarization and polarization angle
-- estimated from satellite borne astrophysical observations, for 
the GRBs, occurring across  the sky.\\
\indent 
This little interesting exercise could have been used to  project out 
a possible 
operating energy range for the space borne detectors, to explore the 
existence of ALPs in the parameter range considered above; had it not 
have faced a  constraint (veto) from the fifth-force experiments due 
to dilatons, that we elaborate below.
%
%

\begin{figure}[ht!]
\begin{minipage}[b]{.50\textwidth}  
\includegraphics[width=1\linewidth]{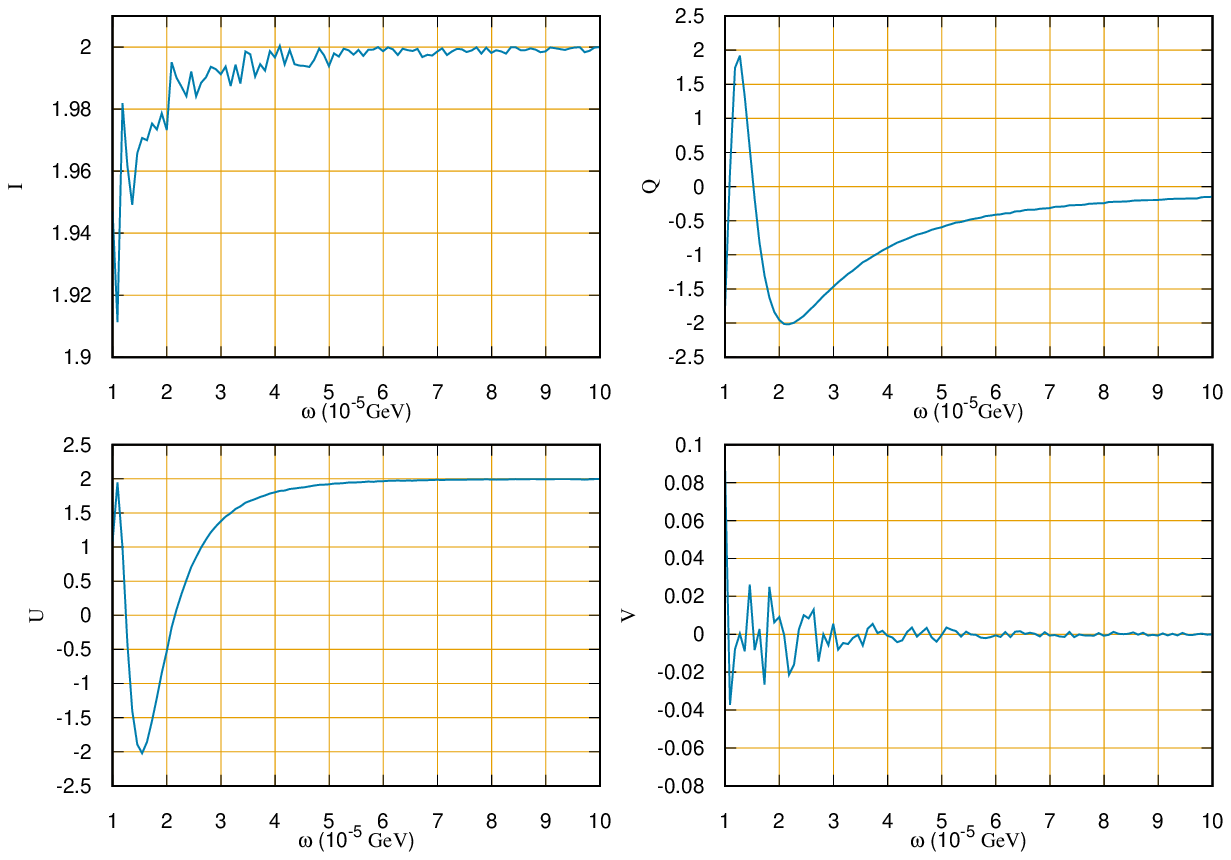}
 \end{minipage}
\begin{minipage}[b]{.50\textwidth}  
\includegraphics[width=1\linewidth]{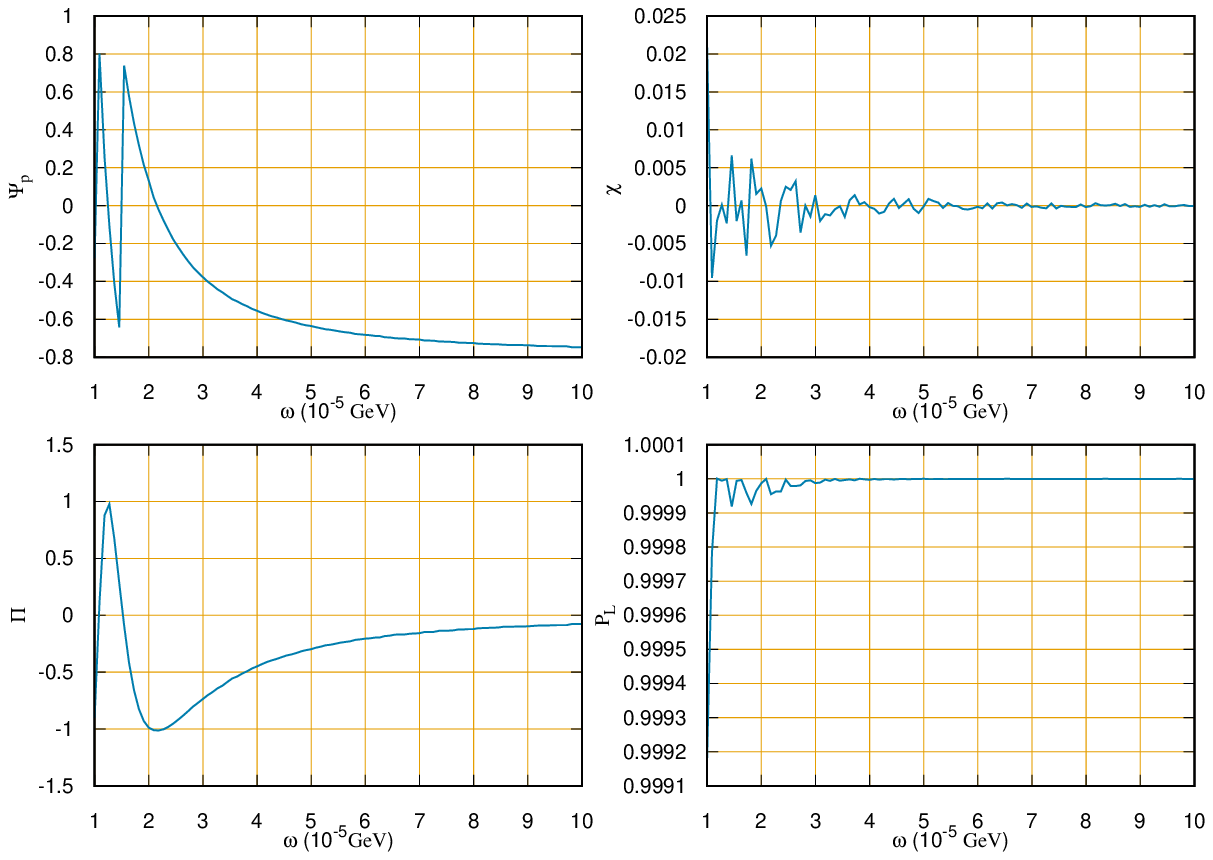}
\end{minipage}
\caption{ (In left panel) plots of Stokes parameters (${\bf I, Q, U, V}$) vs energy ($\omega$). (In right panel) plots of polarization angle ( $\Psi_{p}$, ellipticity angle ($\chi$), degree of polarization ($\Pi$) and linear polarization ($P_{L}$) vs energy ($\omega$).  Parameters considered here i.e., $m_{\phi} = 1.0 \times 10^{-11}$ GeV and $\omega_p = 3.7 \times 10^{-10}$ GeV, respecting fifth force constraints.}
\label{f3}
\end{figure}

\subsection{Fifth force constraints}

\noindent
The scale symmetry, thats being investigated in this work, 
is  an approximate one. This is due to the presence of 
massive dilatons. In field theory context, 
it can happen through the incorporation of higher dimensional 
conformal symmetry breaking operators  (having scaling 
dimension other than four ) or by breaking the symmetry spontaneously
as was done in \cite{salam}.\\
\indent
And in unified theories of extra dimension, this breaking may be
due to nonzero curvature associated with the extra dimension. 
Whatever may be the origin, their existence would modify, the 
predictions of gravitational force of  Einstein gravity with 
dilatonic fifth force. Experiments  performed with torsion-balance,
to find the range of dilatonic fifth force,  puts an upper bound 
on dilaton mass to be $\sim$ $O$($10^{-11}$) GeV. 
Hence,  one needs to consider the value of $m_{\phi} \sim 10^{-11}$ GeV, 
for any realistic search for $\phi$.\\
\noindent
Although we had seen earlier, that, the change in the spectro-polarimetric 
observables with energy ($\omega$) were marginal  when the mass of 
$m_{\phi}$ and plasma frequency $\omega_{p}$ were of the order of 
$\sim 10^{-15}$ GeV; but due to  the reasons explained above those
results are unrealistic.\\ 
\indent
In the light of this, (i.e., for $ m_{\phi} \ge  10^{-2}$ eV) , identifying
the favourable regions in the parameter variables (i.e., estimates of
 $\omega_p$ $eB$ and $g_{\phi\gamma\gamma}$ etc.)  over which the signals 
would remain stabilised--is a time consuming 
task. A way of bypassing this apparent difficulty, is by scaling these 
variables with $ m_{\phi}$ and finding the magnitude of the 
changed variables, such that, the scaled ones remain constant. These 
new modified values are expected to provide a suitable parameter space, 
that one is searching for. We had performed this scaling exercise. And found
the following  modified values for the variables $\omega_p=3.7 \times 
10^{-10}$ GeV, $eB=1.95 \times 10^{-16}$GeV${^2}$ and $g_{\phi\gamma\gamma} 
= 1.0\times10^{-14}$ GeV, rest being the same. The results are provided 
in figure [\ref{f3}]. The numerical exercise confirms the existence of
 of stable signals in the energy range $ 1.0\times 10^{-5} \mbox{~GeV~} <
\omega < 1.0 \times 10^{-4 } $ GeV, for the scaled variables, chosen here.\\ 
\indent
However, in the process, the original parameters, chosen from  the 
generally accepted models of Gamma Ray Bursts, got modified.
So one is left with the option of exploring refined 
models of GRB emission, like synchrotron emission based models or 
hydrodynamic instability based models or streaming instability based 
models etc., as  described in [\cite{waxman}-\cite{Inoue} ].\\   

\section{Discussion}
For obvious reasons, detecting X-ray or gamma- ray signals from earth
surface is difficult. Hence space borne detectors(e.g., GAP) have been 
developed and used to detect the same \cite{yonetokuI}.  They happen
to be sensitive to a wide energy range;  for instance GAP is sensitive 
to an energy range 70 - 300 KeV. The Ranaty High Energy Solar Spectrometer 
Imager (RHESSI) \cite{RHESSI} had an operational window of 0.15 -2 MeV,
the same for SWIFT satellite \cite{SWIFT} happen to be 300 KeV-10 MeV etc.
On the other hand our numerical estimates show that, the signals may 
undergo huge amount of oscillations in these operational windows. A crucial 
assumption that goes in the detection of polarization signal is additivity 
of the Stokes Parameters over an energy range. And the difference in the 
polarization angles of the GRB EM beam,  at two ends of the band, to be less 
than ninety degree, i.e., \cite{yonetokuII}
\begin{eqnarray}
|\Delta(E_2,z)  - \Delta(E_1,z)|  \le \frac{\pi}{2}. 
\label{angle-bound}
\end{eqnarray}
\noindent
 This may not always be the case, as our numerical 
estimates show. Since many of the polarization data-- GRB or others--  
are used to extract estimates about $g_{\phi\gamma\gamma}$ and $m_{\phi}$, 
assuming equation (\ref{angle-bound}) to be valid, when in principle, that 
may not be the case; therefore those estimates of the coupling constant 
and dilaton mass become  questionable.\\

\section{Conclusion}
\noindent
To conclude, in this work we have analysed the mixing pattern
of dilatons with photons in a magnetized medium. Our analysis
establishes that the mixing matrix  for $\gamma\phi$ system is $3\times3$ not $2\times2$.
We also found that, in a magnetized medium, the longitudinal 
DOF associated with the photon, doesn't get excited, 
by the dilatons; the same happens only  for axions, as was rightly pointed out in \cite{wilczek}.
Using analytical techniques we have solved for the equations of 
motion of this system exactly. Following that we have estimated
the strength of the  polarimetric signals that is expected for a
typical GRB geometry.\\
\indent
 What we find is, there are regions in the parameter space over which, 
the signals are stable for some  energy range, and there are regions 
for which the same may not be true ( for instance see  figure [\ref{f2}], where the signal 
($\Psi_{p}$) is seen to stable over photon energy range 
$1.0 \times 10^{-6} $GeV$<\omega < 1\times 10^{-5} $GeV$ $).\\ 
\indent
Given the fact that there are quite a few proposed satellite borne 
experiments in line  [e.g. \cite{APEX} \cite{IXPE} \cite {Hooman} \cite{ASTRO-2020} 
\cite{e-ASTROGAM}], one needs to be careful while designing the detectors 
for them. The operational window of the energy range over which these 
detectors should work-- may be decided after taking into account the lessons 
those emerged from the investigations like ours,  in the respective simulations those usually considered for the purpose of detectors design. \\
\indent
Further more, the  estimates of other observables like the spectral evolutions 
over energy, the magnitude of the fluence, the intensity of the spectrum in 
each polarization channel  etc.,  those usually  emerges from  an analysis 
like ours, should also be taken into account -- while designing the detectors 
( i.e, their energy sensitivity, operational energy range,  dead time 
fixation  and  physical dimension etc.).
\noindent
Once these are taken into account, it would help in getting better quality  
data -- for understanding the nature of the  ALPs and their parameters.\\

\section{ Note} 
  For more details on
discrete symmetries of different  mediums
and their  effects on propagation of EM signals, one can go through  \cite{palash-amj} and  \cite{palash2020}.

\section{Acknowledgment}
The authors would like to thank Prof. Georg G. Raffelt of Max-Planck-Institut fur Physik for being kind enough to going through the manuscript and encouragements.




\section{ Appendix }

 \begin{center}
{\bf { Polarimetry with magnetized media}}
\end{center}

\numberwithin{equation}{section}
\renewcommand{\theequation}{A\arabic{equation}}

In this part of the work, we discuss the polarization effects on an electromagnetic field due to magnetized media.
The  description of  photon propagation in a magnetized media are provided by the set of equations describing   the evolution of  the respective degrees of freedom, one of them having plane of polarization along the magnetic field called  $A_{\parallel}$ and the other one having plane of polarization orthogonal to the same. This set is given by

\begin{eqnarray}
  (k^{2}-\Pi^{2}_{T})A_{\parallel}(\omega,z)+iFA_{\perp}(\omega,z)&=&0,
  \label{mixeqn1}\\
  (k^{2}-\Pi^{2}_{T})A_{\perp}(\omega,z)-iFA_{\parallel}(\omega,z)&=&0.
  \label{mixeqn2}
\end{eqnarray}

\noindent
In  equations (\ref{mixeqn1}) and (\ref{mixeqn2}), the strength of magnetization is carried by  the variable  $F$ and the same is given by  $F=\frac{\omega^{2}_{p}eB\cos\theta}{\omega m_{e}}$. Plasma frequency is denoted by  $\omega_{p}$, $\omega$ is the energy of photon, $eB$ is the magnetic field strength. Defining, $\Omega^{2}= (\frac{\omega^{2}_{p}}{2\omega}-\omega)$. The  eqns. (\ref{mixeqn1}) and (\ref{mixeqn2}) can further be written terms of a mixing matrix ${\bf M}$ causing mixing between the components of polarization $A_{\parallel}$ and $A_{\perp}$, in the following fashion,

\begin{eqnarray}
i\partial_{z} \left[ \left[ \begin{matrix}
1 & 0  \\
 0 & 1\\
\end{matrix}\right] +
\left[ \begin{matrix}
\Omega^{2} & -\frac{i F}{2\omega}  \\
  -\frac{i F}{2\omega} & \Omega^{2} \\
\end{matrix}\right]\right] 
\left[\begin{matrix}
  A_{\parallel}(\omega,z)\\
  A_{\perp}(\omega,z)  \\
\end{matrix}\right]=0.
\label{Mat2}
\end{eqnarray}


\noindent
To find the solutions of $A_{\parallel}$ and $A_{\perp}$, we need eigen values and the eigen vectors for matrix ${\bf M}$. Using the eigen vectors and eigen values $\lambda_{+}$ and $\lambda_{-}$, the unitary matrices turn out to be,

\begin{eqnarray}
 {\tilde U} =\frac{1}{\sqrt{2}}\left[ \begin{matrix}
    i & 1  \\
    \\
  1 & i \\
   \end{matrix}\right]
 {\mbox {~ and,~} }
{\tilde U}^{\dagger}=\frac{1}{\sqrt{2}}\left[ \begin{matrix}
    -i & 1  \\
    \\
  1 & -i \\
   \end{matrix}\right].
\label{Unimat}
\end{eqnarray}

\noindent
Using them, we finally arrive at the solutions of $A_{\parallel}$ and $A_{\perp}$ in terms of the initial conditions $A_{\parallel}(\omega,0)$ $A_{\perp}(\omega,0)$
the solutions are,

\begin{eqnarray}
A_{\parallel}(\omega,z)&=&\left[  \cos(\Omega^{2}z)\cos\left(\frac{Fz}{2\omega}\right)A_{\parallel}(\omega,0)    - \cos(\Omega^{2}z)\sin\left(\frac{Fz}{2\omega}\right)A_{\perp}(\omega,0)\right] \nonumber\\
&+&i\left[\sin(\Omega^{2}z)\cos\left(\frac{Fz}{2\omega}\right)A_{\parallel}(\omega,0) - \sin(\Omega^{2}z)\sin\left(\frac{Fz}{2\omega}\right)A_{\perp}(\omega,0) \right] ,                           
\end{eqnarray}

\begin{eqnarray}
A_{\perp}(\omega,z)&=&\left[  \cos(\Omega^{2}z)\cos\left(\frac{Fz}{2\omega}\right)A_{\perp}(\omega,0)    + \cos(\Omega^{2}z)\sin\left(\frac{Fz}{2\omega}\right)A_{\parallel}(\omega,0)\right] \nonumber\\
&+&i\left[\sin(\Omega^{2}z)\sin\left(\frac{Fz}{2\omega}\right)A_{\parallel}(\omega,0) + \sin(\Omega^{2}z)\cos\left(\frac{Fz}{2\omega}\right)A_{\perp}(\omega,0) \right] .                           
\end{eqnarray}

\noindent
It is easy to check the consistency  of the solutions, by noting that  for $F=0$, there is no mixing between $A_{\parallel}$ and $A_{\perp}$. Now we make use of the solutions to obtain the Stokes parameters $I,Q,U$ and $V$ and following that can evaluate the polarization angle ${\bf \Psi}$ and ellipticity angle $\chi$ from them. The expressions for $I, Q,U $ and $V$ are given as follows,

\begin{eqnarray}
 I&=& A^{2}_{\parallel}(\omega,0) + A^{2}_{\perp}(\omega,0), \\
 Q& =& \cos\left(\frac{Fz}{\omega}\right)\left[  A^{2}_{\parallel}(\omega,0) - A^{2}_{\perp}(\omega,z) \right]-2\sin\left(\frac{Fz}{\omega}\right)A_{\perp}(\omega,0)A_{\parallel}(\omega,0),\\
 U &=& \sin\left(\frac{Fz}{\omega}\right)\left[  A^{2}_{\parallel}(\omega,0) - A^{2}_{\perp}(\omega,0) \right]+2\cos\left(\frac{Fz}{\omega}\right)A_{\perp}(\omega,0)A_{\parallel}(\omega,0).
\end{eqnarray}

\noindent
It turns out that, in this case the Stokes parameter $V$ describing circular polarization is equal to zero. It is to be noted that the Stokes parameter $I$ is now independent of the path length $z$, which provides the consistency check of energy conservation of the system. Also, the  rate of rotation of the plane of polarization, with change of path length distance, is proportional to the inverse square of energy of photon, i.e.,

\begin{eqnarray}
  \frac{d\bf{\psi}}{dz} = \frac{\omega^{2}_{p}eB\cos\theta}{2\omega^{2}m_{e}}.
  \end{eqnarray}

\noindent
The $\omega$ dependence of this result matches with the same reported in \cite{ganguly-konar-pal}. Therefore one can state that the effect of magnetized medium alone on the polarimetric signature of light can be predicted by studying the state of circular polarization of light along with energy dependence of rate of rotation of the polarization angle per unit length by studying the system when the magnetic field $\vec{B}$ is along $\vec{k}$. On the other hand when the angle between $\vec{k}$ and $\vec{B}$ is $\frac{\pi}{2}$ then the effect of magnetized media of polarization of the electromagnetic beam vanishes and the variation of polarization with $\omega$ can be found in \cite{JKPS}. Similar studies only for scalars or pseudoscalars have been performed in  \cite{Pankaj-Ralston}, that show different outcome.


\section{Supplementary document}
\numberwithin{equation}{section}

\renewcommand{\theequation}{S\arabic{equation}}

\renewcommand{\thesection}{S\arabic{section}}


\subsection{Discrete symmetry effects on propagation}
We begin this section by defining the notations, used in this paper. We work 
in flat space with the metric tensor defined as  $g^{\mu\nu}=g_{\mu\nu}=
\mbox{diag}(+1,-1,-1,-1 )$. The contravariant  and covariant four-vectors are 
defined as follows,
\begin{eqnarray}
x^{\alpha}=(t, \vec{x}), \mbox{~~~~~~} x_{\alpha}=g_{\alpha\beta}x^{\beta} = 
(t,-\vec{x})^{T}.
\end{eqnarray} 
\noindent
In our notation, the Greek indices are supposed to take values from 0 to 3 and 
the Latin ones between 1 to 3.\\
\indent
The orthochronous improper  Lorentz transformation, also called parity 
inversion transformation and non-orthochronous improper Lorentz 
transformation, also called the time reversal transformation, are defined 
by the matrices,
\begin{eqnarray}
{\cal{P}}^{\mu}_{~~\nu} = {\left({\cal{P}}^{-1}\right)}^{\mu}_{~~\nu} = 
\begin{pmatrix}
1 & 0 & 0 & 0 \\
0 & -1 & 0 & 0 \\
0 & 0 & -1 & 0 \\ 
0 & 0 & 0 & -1 \\
\end{pmatrix}
\mbox{~~~~and~~~~}
{\cal{T}}^{\mu}_{~~\nu} = {\left({\cal{T}}^{-1}\right)}^{\mu}_{~~\nu} = 
\begin{pmatrix}
-1 & 0 & 0 & 0 \\
0 & 1 & 0 & 0 \\
0 & 0 & 1 & 0 \\ 
0 & 0 & 0 & 1 \\
\end{pmatrix}.
\label{PT}
\end{eqnarray}
\noindent
The corresponding unitary operators for parity and time reversal 
transformations, are identified as $U(P)\equiv {\bf P}$ and $U(T)\equiv  {\bf T}$. 
Their action on any scalar $\Phi(x)$, generates the following
transformation, 
\begin{eqnarray}
{\bf P} \Phi(x) {\bf P}^{-1} = \Phi({\cal{P}}^{-1}x) \mbox{~~and~~} {\bf T} \Phi(x) {\bf T}^{-1}
= \Phi({\cal{T}}^{-1}x).
\label{defining-transf-PT-}
\end{eqnarray}
\noindent
Since ${\bf P}$ and ${\bf T}$ are their own {\em inverse}, therefore the right hand 
side of equation (\ref{defining-transf-PT-}), can also be represented in the
following way, 
\begin{eqnarray}
{\bf P} \Phi(x) {\bf P}^{-1} = \Phi({\cal{P}}x) \mbox{~~and~~} {\bf T} \Phi(x) {\bf T}^{-1}
= \Phi({\cal{T}}x),
\label{defining-transf-PT+}
\end{eqnarray}
for time reversal and parity inversion transformations.\\

\noindent
The  complex number $i = \sqrt{-1}$, under time reversal undergoes 
anti-unitary transformation,  so that, the Hamiltonian ($H$) a self 
adjoint operator, ( also represented by the zeroth component of the 
momentum four vector, $k^{0} $), retains the right sign under time 
(${\bf T}$) reversal. However there are no such restriction for {\bf P} 
and {\bf T}. So $i$ would remain inert, under them. Hence the 
transformation laws for $i$ are,
\begin{eqnarray}
{\bf T}(i)  {\bf T}^{-1} &=& -i,
\label{i-T} \\
{\bf P} (i) {\bf P}^{-1}&=& +i,
\label{i-P} \\
{\bf C} (i) {\bf C}^{-1}&=&  +i.
\label{i-cpt}
\end{eqnarray}

\indent
Lastly, the totally anti-symmetric Levi-Civita tensor, defined as
$\epsilon^{0123}=-\epsilon_{0123}=1$; picks up a $-$ve sign under both 
parity $({\bf P})$ and time reversal (${\bf T}$) transformations. That is, under 
parity transformation, it transforms as, 
${\bf P}\epsilon^{\alpha\beta\gamma\sigma} {\bf P}^{-1}= - \epsilon^{\alpha\beta\gamma\sigma}
= \epsilon_{\alpha\beta\gamma\sigma}$
and under time reversal, it transforms as,
 ${\bf T}\epsilon^{\alpha\beta\gamma\sigma} {\bf T}^{-1}= - \epsilon^{\alpha\beta\gamma\sigma}
=\epsilon_{\alpha\beta\gamma\sigma}$. 
In compact notation one can represent both of them as,
\begin{eqnarray}
{\bf T}\epsilon^{\alpha\beta\gamma\sigma} {\bf T}^{-1}= \epsilon_{\alpha\beta\gamma\sigma},\\
{\bf P}\epsilon^{\alpha\beta\gamma\sigma} {\bf P}^{-1}= \epsilon_{\alpha\beta\gamma\sigma}, \\
{\bf C}\epsilon^{\alpha\beta\gamma\sigma} {\bf C}^{-1}= \epsilon^{\alpha\beta\gamma\sigma}.
\label{PT-transf}
\end{eqnarray}
The last line follows from, the fact that, the net  charge 
associated with any system remains the same under a duality 
transformation.
\subsection{Discrete transformation properties of the fundamental 
vectors.}

\noindent
Next we come to discuss the transformation properties of space time four vector
$x^{\mu}$. Under time  reversal ({\bf T}), space inversion 
({\bf P})  and charge conjugation ({\bf C }) transformations, 
the four vector  $x^{\mu}$  transforms as,
\begin{eqnarray}
{\bf T} x^{\mu}{\bf T}^{-1} &=& (-t, \vec{x}) = -x_{\mu},
\label{x1}\\
{\bf P}x^{\mu}{\bf P}^{-1} &=& (t, -\vec{x}) = +x_{\mu},
\label{x2}\\
{\bf C}x^{\mu}{\bf C}^{-1} &=& (t, \hphantom{-}\vec{x}) = +x^{\mu}.
\label{x3}
\end {eqnarray}
\noindent
The last equation  equation (\ref{x3}) follows from the fact that, under 
charge conjugation the  space-time  4-vectors undergo no change.\\ 
\indent
In the light of equations ((\ref{x1}) -- (\ref{x3})), we can discuss the 
transformation rules of the momentum 4-vectors associated with the 
system. The same by the virtue of being  observables, are supposed to 
be  self-adjoint operators. Therefore following the rules of relativistic 
quantum theory, they can be represented by; 
\begin{eqnarray}
k^{\mu} \!\!\!\!\!&\equiv& \!\!\!  i\partial^{\mu}  = i(\partial^{0}, \partial^{j})  
= i\frac{\partial}{\partial x_{\mu}} = i(\partial_0, 
-\mathbf{\nabla} ),
\label{k1}
\\
k_{\mu}\!\!\!\!\!&\equiv& \!\!\! i\partial_{\mu} = i(\partial_{0}, \partial_{j}) = i\frac{\partial}{\partial x^{\mu}} =
 i(\partial_0, 
\hphantom{-}\mathbf{\nabla} ).
\label{km2}
\end{eqnarray} 
\noindent
Using equations ((\ref{i-T}) -- (\ref{i-cpt})), in conjugation with 
 ((\ref {x1}) -- (\ref {x3})) in ((\ref {k1}) and (\ref{km2})), 
its easy to see that, under  the operation of {\bf T} reversal 
transformation, $k^{\mu}$ transforms as; 
${\bf T} k^{\mu}{\bf T}^{-1} = -i(-\partial^{0}, \partial^{j})  = i(\partial^{0},- \partial^{j}) = i\partial_{\mu}=k_{\mu}$.  Similarly, under {\bf P}, it transforms as 
${\bf P} k^{\mu}{\bf P}^{-1} = i(\partial^{0}, -\partial^{j})  
= i(\partial_{0},\partial_{j}) = i\partial_{\mu}= k_{\mu}$. However 
$k^{\mu}$ remains the same under {\bf C} transformation.  Hence, the 
transformation laws of $k^{\mu}$ individually under {\bf T,~P } and {\bf C}
are given by,
\begin{eqnarray}
{\bf T} k^{\mu}{\bf T}^{-1} &=& k_{\mu}.\\
{\bf P} k^{\mu}{\bf P}^{-1}&=&  k_{\mu}, \\
{\bf C} k^{\mu}{\bf C}^{-1}&=&  k^{\mu},
\end{eqnarray}

\noindent
Finally, the centre of mass four velocity of the medium, defined as, $u^{\mu}=\frac{dx^{\mu}}{d\tau}$,( when $d\tau = \sqrt {dt^{2} - dx^{2}}$ is the differential proper-time interval ), have the following transformation properties 
under time reversal, parity and charge conjugation transformations:
\begin{eqnarray}
{\bf T} u^{\mu}{\bf T}^{-1} &=& (-u^{0}, u^{j})  = -u_{\mu},\\
\label{t-u}
{\bf P} u^{\mu}{\bf P}^{-1} &=& (u^{0}, -u^{j}) = +u_{\mu},\\
\label{p-u}
{\bf C}u^{\mu} {\bf C }^{-1}&=& (u^{0}, -u^{j}) =-u^{\mu}.
\label{c-u} 
\end{eqnarray}
 
\noindent
The last property (\ref{c-u}) follows from the observation, that the statistical part of the thermal 
propagator in real time thermal QFT should remain invariant under the operation of
charge conjugation. 
The factor that carry the information about temperature and chemical potential of a fermionic quantum system 
according to quantum statistical field theory, is given by,
\begin{eqnarray}
\eta_{T}(p.u)=\left[ \Theta(p.u) n_{F}(p.u, \mu, \beta)+    \Theta(-p.u) n_{F}(-p.u,- \mu, \beta)  \right],
\label{thermal}
\end{eqnarray}
with $ \Theta(p.u)$ the step function and $n_{F}(p.u,\mu,\beta)= \frac{1}{e^{ \left(\frac{p.u - \mu}{\beta}  \right)  }  +1} $ the  Fermi distribution function, and 
$\beta$ the inverse of temperature. One can see that under the operation  of charge conjugation ({\bf C}) transformation, if $\mu \to -\mu$,
then in order to have the statistical factor $\eta_{T}(p.u)$ inert under the same, 
one should have 
$u^{\mu}  \to -u^{\mu}$ 
\cite{ganguly-konar-pal}. As explained there (\!\!  \cite{ganguly-konar-pal}), the same in the rest frame of the medium,
is given by $ u^{\mu}= \left( 1, 0,0,0 \right)$; for the remaining part of this paper we shall 
assume this to be true.\\       

\subsection{Discrete transformation properties  of the fields  }
Here we note down the transformation properties  of four-vector potential $A^{\mu}(x^{\alpha})$, followed by field strength tensor $F^{\mu\nu}(x^{\alpha})$ under the set of discrete transformations (i.e., {\bf C, P }and {\bf T} ). 
The transformation properties of $A^{\mu}(x^{\alpha})$ following \cite{Itzykson-Zuber}, under  
parity {\bf P}, time reversal {\bf T} and charge conjugation {\bf C} are:  
\begin{eqnarray}
{\bf T} A^{\mu}(x^{\alpha}){\bf T}^{-1} &=& +A_{\mu}(-x_{\alpha}), 
\label{AT}\\
{\bf P} A^{\mu}(x^{\alpha}){\bf P}^{-1} &=&+A_{\mu}(x_{\alpha}),
\label{Ap}\\
{\bf C }A^{\mu}(x^{\alpha}) {\bf C}^{-1} &=& -A^{\mu}(x^{\alpha}).
\label{Ac}
\end{eqnarray} 

\noindent
Using the equations (\ref{AT}) to (\ref{Ac}) and the transformation  properties of ${\partial_{\nu}}$ under 
{\bf C}, {\bf P} and {\bf T}, the transformation properties  of the field strength tensor $F^{\mu\nu} (x)$, under 
them can be found. And they happen to be:\\ 
\begin{eqnarray}
{\bf T} F^{\mu\nu}(x^{\alpha}){\bf T}^{-1} &=& -F_{\mu\nu}(-x_{\alpha}),
\label{TFmunu}\\
{\bf P} F^{\mu\nu}(x^{\alpha}){\bf P}^{-1} &=& +F_{\mu\nu}(x_{\alpha}),
\label{PFmunu}\\
{\bf C} F^{\mu\nu}(x^{\alpha}) {\bf C}^{-1} &=& -F^{\mu\nu}(x^{\alpha}).
\label{CFmunu}
\end{eqnarray}
Finally, the relation between the electric or magnetic field components and  the components of the field strength tensor,
$F^{\mu\nu}$ are given by,

\begin{eqnarray}
F^{\mu\nu} = 
\begin{pmatrix}
0 & -E_{x} & -E_{y} &  -E_{z}  \\
 E_{x} & 0 &  -B_{z} &  B_{y} \\
 E_{y} &  B_{z} & 0 &  -B_{x} \\ 
 E_{z} &  -B_{y}&  B_{x} & 0 \\
\end{pmatrix}
\mbox{~~~~and~~~~}
F_{\mu\nu}= g_{\mu\alpha}F^{\alpha\beta}g_{\beta\nu}= 
\begin{pmatrix}
0 & E_{x} & E_{y} &  E_{z}  \\
 -E_{x} & 0 &  -B_{z} &  B_{y} \\
 -E_{y} &  B_{z} & 0 &  -B_{x} \\ 
 -E_{z} &  -B_{y}&  B_{x} & 0 \\
\end{pmatrix}.
\label{PT2}
\end{eqnarray}

\noindent
Using the relation given by (\ref{PT2}) and ((\ref{TFmunu})-(\ref{CFmunu})) one can  
establish the {\bf C, P} and {\bf T} transformation properties of the electric  $ E$ 
and magnetic $B$ fields. We have performed the same and our transformation properties
obtained this way matches with that of Jackson \cite{jack}\\

\subsubsection{ {\bf{C}}, {\bf{P}} and {\bf{T}} transformations properties of the basis vectors  :}

\noindent
We next find out the  $\bf{C,P,\mbox{ and } T}$ transformation properties
of the basis vectors. We start with vector $b^{(1)\mu}$, given by
$ b^{(1)\mu}=k_{\nu} F^{\mu\nu} $. Using the transformation properties
of $ k_{\nu}$ and $F^{\mu\nu}$ under {\bf P}, {\bf T} and {\bf C}, one 
can show that: 
\begin{eqnarray}
{\bf{T}}b^{(1)\mu}{\bf{T}}^{-1} &=& -b^{(1)}_{\mu}, 
\label{t-b1i}\\
{\bf{P}} b^{(1)\mu}  {\bf{P}}^{-1} &=&+ b^{(1)}_{\mu}, 
\label{p-b1i}\\
{\bf{C}} b^{(1)\mu} {\bf{C}}^{-1}&=&-b^{(1)\mu} .
\label{c-b1i} 
\end{eqnarray}

\noindent
Similarly using the definition of the 4-vector,  $b^{(2)\mu}$, as $b^{(2)\mu}= \frac{1}{2}k_{\beta}
\bar{F}_{\lambda\rho}\epsilon^{\beta\lambda\rho\mu}$ one can show the transformations
of the same under {\bf T},{\bf P} and {\bf C}. They happen to be, 
\begin{eqnarray}
{\bf T} b^{(2)\mu} {\bf T}^{-1} &=& -b^{(2)}_{\mu},
\label{b2T}\\
{\bf P} b^{(2)\mu} {\bf P}^{-1} &=& +b^{(2)}_{\mu},
\label{b2P}\\
{\bf C} b^{(2)\mu} {\bf C}^{-1} &=& -b^{(2)\mu}.
\label{b2C}
\end{eqnarray}

\noindent
Next we are left with estimating the same transformation rules for four vectors $I^{\mu}$ and $\tilde{u}^{\mu}$.
But before we perform that, we need to have the transformations of, $b^{(2)}\cdot{\tilde u}$ and $k \cdot u$ 
under {\bf{C}}, {\bf{P}} and {\bf{T}}. The dot products in  $b^{(2)}\cdot{\tilde u}$ and $k \cdot u$  are
to be understood as dot products in 4-dimensions. The transformations  for $b^{(2)}\cdot{\tilde u}$  are,
\begin{eqnarray}
{\bf{T}} \left( \tilde{u}\cdot b^{(2)}\right) {\bf{T}}^{-1} &=&  \left( \tilde{u}\cdot b^{(2)} \right),   \\
{\bf{P}} \left( \tilde{u}\cdot b^{(2)}\right) {\bf{P}}^{-1} &=&  \left( \tilde{u}\cdot b^{(2)} \right),  \\
{\bf{C}} \left( \tilde{u}\cdot b^{(2)}\right) {\bf{C}}^{-1} &=&  \left( \tilde{u}\cdot b^{(2)} \right). 
\label{b2utilde}
\end{eqnarray}
\noindent
Similarly, follows the transformation  laws  for the four dimensional dot product $k \cdot u$, from the individual transformation 
laws of $k^{\mu}$ and $u_{\mu}$, and they are, 
\begin{eqnarray}
{\bf{T}} \left( u\cdot k \right) {\bf{T}}^{-1} &=& - \left( u
\cdot k \right),  \\
{\bf{P}} \left( u\cdot k \right) {\bf{P}}^{-1} &=& + \left( u
\cdot k \right), \\
{\bf{C}} \left( u\cdot k \right) {\bf{C}}^{-1} &=&  -\left( u
\cdot k \right). 
\label{utildek}
\end{eqnarray}
Recalling, the four vector $I^\nu$, to be given by
\begin{equation}
 I^{\nu} = \left(b^{(2)^\nu} - \frac{(\tilde{u}^\mu b^{(2)}_\mu)}{\tilde{u}^2}\tilde{u}^\nu \right);\nonumber
 \label{para-pol-vec-2}
 \end{equation} 
the transformation laws for, $ I^{\mu} $, using the transformation laws for 
equation (\ref{b2utilde}) turns out to be,
\begin{eqnarray}
{\bf{T}} I^{\mu} {\bf{T}}^{-1} &=& - I_{\mu}, \\
{\bf{P}} I^{\mu} {\bf{P}}^{-1} &=& + I_{\mu},  \\
{\bf{C}} I^{\mu} {\bf{C}}^{-1} &=& - I^{\mu}.   
\label{capI}
\end{eqnarray}
Lastly the same for $\tilde{u}^{\mu}$, using equation (\ref{utildek}) are found
to be, 
\begin{eqnarray}
{\bf{T}} {\tilde{u}}^{\mu} {\bf{T}}^{-1} &=& - {\tilde{u}}_{\mu},\\
{\bf{P}} {\tilde{u}}^{\mu} {\bf{P}}^{-1} &=& + {\tilde{u}}_{\mu}, \\
{\bf{C}} {\tilde{u}}^{\mu} {\bf{C}}^{-1} &=& - {\tilde{u}}^{\mu}.
\label{tildeu}
\end{eqnarray}

\subsection{Discrete transformation properties: EM form-factors}
\noindent
The EM form-factors associated with the gauge potential, $A_{\mu}$,
are $A_{\parallel}$,  $A_{\perp}$,  $A_{L}$ would have their own transformation laws
under {\bf  C, P} and {\bf T}. In the next few lines we would outline the
detail of finding their transformation rule. The basic principle lies with
finding out the transformation rules for the time-like and the space like 
components  of  $A_{\mu}$, following from equations (\ref{AT}) to (\ref{Ac}).
We recall the following transformations, that follows from,  
equations (\ref{AT}) to (\ref{Ac}),
\begin{eqnarray}
{\bf T}A^{0}{\bf T}^{-1}= +A^{0}, \,\,\,{\bf P}A^{0}{\bf P}^{-1}= +A^{0}, \,\,\,{\bf C}A^{0}{\bf C}^{-1}= - A^{0}, \nonumber \\
{\bf T}A^{i}{\bf T}^{-1}= -A^{i}, \,\,\, {\bf P}A^{i}{\bf P}^{-1}= -A^{i},\,\,\,\,~{\bf C}A^{i}{\bf C}^{-1}= - A^{i}.
\label{CPT-condns}
\end{eqnarray}
Next we can express the time-like and the space-like parts of the  
gauge potentials in terms of  the EM form-factor from the
 definition of the gauge potential;  
\begin{eqnarray}
A^{\alpha}(k)= A_{\parallel}(k)\rm{N}_1 b^{(1)\alpha}+ A_{\perp}(k) \rm{N}_2I^{\alpha}
+ A_{L}(k)\rm{N}_L\tilde{u}^{\alpha}.
\label{gauge-p}
\end{eqnarray}
\noindent
They turn out to be,
\begin{eqnarray}
A^{0}&=&N_1 b^{(1)0} A_{\parallel} + N_{2}I^{0} A_{\perp}+ N_{L} \tilde{u}^{0}A_{L},
\label{a-time}\\
A^{i}&=&N_1 b^{(1)i} A_{\parallel} + N_{2}I^{i} A_{\perp}+ N_{L} \tilde{u}^{i}A_{L}. 
\label{a-space}
\end{eqnarray}
\noindent
If we use the  relations given by equation (\ref{CPT-condns}) on equations 
(\ref{a-time}) and (\ref{a-space}), we would arrive at,
\begin{eqnarray}
{\bf T} A_{\parallel} {\bf T}^{-1} =  -A_{\parallel}, \,\,{\bf T} A_{\perp} {\bf T}^{-1} =  -A_{\perp} , \,\,{\bf T} A_{L} {\bf T}^{-1} = -A_{L},  \\
\label{timelike}
{\bf P} A_{\parallel} {\bf P}^{-1} =  +A_{\parallel}, \,\,{\bf P} A_{\perp} {\bf P}^{-1} =  + A_{\perp}, \,\,{\bf P} A_{L} {\bf P}^{-1} =  +A_{L}, \\
\label{spacelike}
{\bf C} A_{\parallel} {\bf C}^{-1} =  +A_{\parallel},\,\, {\bf C} A_{\perp} {\bf C}^{-1} =  +A_{\perp}, \,\,{\bf C} A_{L} {\bf C}^{-1} =  +A_{L}. 
\label{chargelike}
\end{eqnarray}
\noindent 
once we take the transformation laws of the basis vectors under {\bf T, P } and {\bf C} into account.

\subsection{The photon polarization tensor in magnetized media.}

The photon polarization tensor, $\Pi_{\mu\nu}^{p}(k)$, in a magnetized medium
to $O(eB)$, can be expressed parametrically in the following form,\\
\begin{eqnarray}
\Pi^{p}_{\mu\nu}(k)& = &\Pi^{p}(k) \frac{  i}{\sqrt{(ku)^{2}-k^{2}}}
\epsilon_{\mu\nu\beta_{\parallel}\delta}u^{{\tilde\beta}_{\parallel}} k^{\delta}.
\label{pi_p}
\end{eqnarray}
\noindent
The notations being, in the Levi-Civita tensor $\epsilon_{\mu\nu\beta_{\parallel}\delta}$, the subscripted index ${\beta_{\parallel}}$ is allowed to take only two
values, either 0 or 3; although the subscripted index 
${\tilde\beta}_{\parallel}$  is also allowed to take only
those two numerical values,{\it but there is a  difference}; the difference being, when  ${\beta_{\parallel}}= 0$  then ${\tilde\beta}_{\parallel}=3$ and when ${\beta_{\parallel}}= 3$  then 
${\tilde\beta}_{\parallel}=0$.\\
\indent
 This leaves us with fixing the functional form 
of the  argument of the scalar EM form-factor $\Pi^{p}(k)$. 
The functional form of the arguments have to be Lorentz scalar.
Though, in principle these Lorentz scalars can be of the form: $u.k$ or $k^2$ etc., 
however their power and functional-structure  would be decided on the basis 
of some general Quantum Field Theoretic  arguments; that a two point function is supposed to 
follow.\\    
\indent
 One of these formal  Quantum Field Theoretic arguments is, that the partition function  
or S matrix should be unitary: this requirement dictates that the two point
function (photon polarization tensor) must be hermitian i.e., 
\begin{eqnarray}\Pi^{p}_{\mu\nu}(k) = \Pi^{p*}_{\nu\mu}(k).
\label{hermiticity}
\end{eqnarray}  
\noindent
It should obey Bose symmetry, i.e., 
\begin{eqnarray}
\Pi^{p}_{\mu\nu}(k) = \Pi^{p}_{\nu\mu}(-k).
\label{Bose-symmetry}
\end{eqnarray}\\
\noindent
Additionally, since the same is being evaluated for a theory that is 
{\bf CPT} invariant, the polarization tensor should also remain invariant under 
the combined operation of {\bf CPT} transformation, as well as-- 
invariant separately-- under  {\bf C } and {\bf PT}. That is:\\
\begin{eqnarray}
\!\!\!\!\!\mbox{\bf (CPT)} \Pi^{p}_{\mu\nu}(k) \mbox{\bf (CPT)$^{-1}$} = \Pi^{p}_{\mu\nu}(k), 
\,\,
\mbox{\bf C} \Pi^{p}_{\mu\nu}(k) \mbox{\bf C$^{-1}$} = \Pi^{p}_{\mu\nu}(k) 
\mbox{~~~\&~~~}
\mbox{\bf (PT)} \Pi^{p}_{\mu\nu}(k) \mbox{\bf (PT)$^{-1}$} = \Pi^{p}_{\mu\nu}(k)
\end{eqnarray}\\
\noindent 
Fulfilment of the condition of {\it hermiticity}  
given by equation (\ref{hermiticity}), is satisfied due to the presence
of the Levi-Civita symbol along with the multiplicative factor of $i$ $
(i.e.,\sqrt{(-1)})$, below the pair production threshold.\\
\indent
 Similarly, the condition of Bose symmetry given by equation (\ref{Bose-symmetry}), is fulfilled
due to the presence of the same  Levi-Civita symbol multiplied by four 
vector $k$. This however puts strong restriction on the functional form of 
form-factor $\Pi^{p}(k)$, that is, it just tells us that, the same has to be
even function of the scalars like $k^2$  or  $k.u$ etc.   Evaluation of 
the same from the corresponding one loop Feynman diagram --using perturbation
theory shows -- that, it indeed is a function of $\sqrt{(k.u)^2}=\omega$.\\ 

\indent
The Charge conjugation
symmetry, similarly dictates that: since the tensorial structure is odd under 
${\bf C}$ therefore to maintain overall {\bf C} invariance the form-factor
$\Pi^{p}(k)$ should be either odd in B or odd in chemical potential $\mu$,
but not both. Since  we have evaluated this term to order $eB$ and that too
 as a scalar in $B$, hence it can be utmost be a function of $\sqrt{B^{2}}$. 
Hence it should be  odd in chemical potential. The same is achieved 
automatically, if the statistical factor is composed of the difference
between electron and positron distribution functions 
$(n_{e^{\pm}}(p.u,\mu,\beta))$. That is with the presence of a piece
that represent the difference in the fermionic densities, 
like $(n_{(e^-)} - n_{(e^+)})$.\\

\indent
Having discussed the structure of the form-factor on general grounds, we now 
provide with the expressions of the form-factor $\Pi^{p}(k)$ as obtained from 
analytical evaluation of the same carried out in \cite{ganguly-konar-pal} 
vi-a-vis the same obtained on the ground of the general analysis presented 
above.  The same turns out to be,
\begin{eqnarray}
\Pi^{p}(k) = \underbrace{\frac{\sqrt{(k.u)^2}\sqrt{(eB/m_e)^2}}{\omega^2 - (eB/m_e)^2}
\left(\frac{n_e}{m_e} \right)}= \frac{\omega \omega_B \omega_p^2}
{\omega^2 - \omega_B^2}
\label{pip-ana}
\end{eqnarray}\\
In the expression above, given by equation (\ref{pip-ana}), 
$\omega_B=\frac{eB}{m_e}$ also called the Larmor frequency and $\omega_p$, is 
the plasma frequency, given by $\omega_p= \sqrt{\frac{4\pi \alpha n_{(e^{-})}}{m_e}}$. The expression at the middle of eqn. (\ref{pip-ana})( with an underbrace),
 represents the structure of $\Pi^{p}(k)$, as anticipated on the basis of 
our general arguments. And the term next to it, on the right hand side of the
equal to sign, represents the same obtained after evaluating the 
one-loop-Feynman diagram.
We conclude this section with the observation that, the  EM 
form-factor $\Pi^{p}(k)$, appearing from the self energy correction  due 
to magnetized matter effects, is {\bf C} odd ( due to $\omega^2_p \propto n_{e^{-}}$) but {\bf PT } even.

\subsection{Discrete transformations of EOM}
The equation of motions for scalar-photon ($\phi\gamma$) system are as follows;

\begin{eqnarray}
(k^2-\Pi_T)A_{\parallel} (k)+ i\Pi^{p}(k)N_{1}N_{2}\left[\epsilon_{\mu\nu\delta\beta}
\frac{k^{\beta}}{\mid{k}\mid}u^{\tilde{\delta}_{\parallel}} b^{(1)\mu} I^{\nu}\right] A_{\perp}(k)
&=& \frac{ig_{\phi\gamma\gamma}\phi(k)}{N_1},
\label{apfd1} \\
(k^2-\Pi_T)A_{\perp} (k) - i\Pi^{p}(k)N_{1}N_{2}\left[\epsilon_{\mu\nu\delta\beta}
\frac{k^{\beta}}{\mid{k}\mid}u^{\tilde{\delta}_{\parallel}} b^{(1)\mu} I^{\nu}\right]A_{\parallel}(k)&= &0,
\label{apfd2}\\ 
(k^2 -\Pi_L)A_L(k) &=& 0, 
\label{apfd3}\\
(k^2 - m^2)\phi(k) &=& -\frac{ig_{\phi\gamma\gamma} A_{\parallel}(k)}{N_1}. 
\label{apfd4}
\end{eqnarray}

\noindent
An useful exercise towards establishing the correctness of the set of equations
of motions, is to show that, these equations obey some symmetry condition. The 
symmetry being the {\bf PT} symmetry here. The first step towards establishing 
that, is to neglect  $\Pi^{p}(k)$ in equations (\ref{apfd1} -- \ref{apfd4}). 
One can see that, after this exercise, only equations (\ref{apfd1}) and (\ref{apfd4}) remain coupled and the rest gets decoupled. 
This is expected, because, in the tree level interaction
 Lagrangian $(L_I=\frac{1}{4} \phi {\bar{F}_{\mu\nu}}f^{\mu\nu})$, 
the scalar  $\phi$ has no coupling with either $A_{L}$ or $A_{\perp}$. 
Using the decomposition of the vector potential in terms of the 
basis vectors,
\begin{eqnarray}
A^{\nu}(k)= A_{\parallel}(k)\rm{N}_1 b^{(1)\nu}+ A_{\perp}(k) \rm{N}_2I^{\nu}+ A_{L}(k)
\rm{N}_L\tilde{u}^{\nu} + \rm{N}_k A_{gf}(k)k^{\nu},
\label{gauge-pot2}
\end{eqnarray}
the issues related to coupling of various form factors, can be verified, by substituting (\ref{gauge-pot2})
 in,
\begin{eqnarray}
L_I=\frac{1}{4} \phi {\bar{F}_{\mu\nu}}f^{\mu\nu}.
\label{TL-mag}
\end{eqnarray}
\noindent
That is, substituting (\ref{gauge-pot2}) in the expression of the dynamical 
part of the photon field strength tensor, $f_{\mu\nu}$ in (\ref{TL-mag}).
This also turns out to be true if effective Lagrangian with istropic
self-energy tensor,
$ A_{\mu}\Pi^{\mu\nu}(T,\mu)A_{\nu}$ along with magnetic field induced
tree level Lagrangian given by (\ref{TL-mag}) effects are considered.\\ 

\indent
Now operating {\bf T} or {\bf PT} on both sides of (\ref{apfd1}) and (\ref{apfd4}) one can convince one self ,that the change in sign due to the discrete 
transformations are compensated with-out changing the equation structures.
That is the equations remain invariant.
To proof of  the same when $\Pi^{p}(k)$  is retained, would require 
the transformation properties of the quantities (under {\bf T, C,P}) 
provided below.\\

 \begin{eqnarray}
    ({\bf PT}) A_{\parallel} ({\bf PT})^{-1}  &=& -A_{\parallel}, 
    \label{pt-Aparallel}\\
    ({\bf PT}) A_{\perp} ({\bf PT})^{-1}  &=& -A_{\perp},
    \label{pt-Aperp}\\
    ({\bf PT}) i ({\bf PT})^{-1}  &=& -i,  
    \label{pt-i}\\
    ({\bf PT}) k^{\beta} ({\bf PT})^{-1}  &=& +k^{\beta},
    \label{pt-k}\\
    ({\bf PT}) u^{\delta_{\parallel}}({\bf PT})^{-1}  &=& -u^{\delta_{\parallel}},
    \label{pt-u}\\
    ({\bf PT}) b^{(1)\mu} ({\bf PT})^{-1}  &=&  -b^{(1)\mu},
    \label{pt-b1}\\
    ({\bf PT}) I^{\nu} (PT)^{-1}  &=& -I^{\nu},
    \label{pt-i}\\
    ({\bf PT}) \epsilon_{\mu\nu\delta\beta} ({\bf PT})^{-1}  &=&  \epsilon_{\mu\nu\delta\beta}, 
    \label{pt-i}\\
    ({\bf PT}) \Pi^{p}(k)({\bf PT})^{-1}  &=&  \Pi^{p}(k). 
    \label{pt-eps}
\end{eqnarray}

\noindent
One can see that, if the equations (\ref{apfd1}) -- (\ref{apfd4}), are multiplied 
from left by  {\bf (PT)} and from right by {\bf (PT$)^{-1}$} they 
remain invariant . That establishes that the equations are  {\bf (PT)} 
symmetric.\\
\noindent
The form-factor corresponding to the longitudinal DOF
$A_L$, still remains decoupled and, this DOF of photon
 propagates freely. This happens because, the scalar field 
$\phi$ and$A_L$ has no interaction at the tree level  of the 
interaction Lagrangian. \\
\indent

\indent
Next we come to the structure of the mixing matrix. It should be noted 
that in absence of scalar photon interaction and $\Pi^{p}(k)$, the
normal modes of the system are given by the three DOF, 
the transverse DOF $A_{\parallel}$, $A_{\perp}$ and the longitudinal DOF $A_L$.
Such a system will be described by a $3\times3$ matrix, where the only the
diagonal elements are nonzero. Upon the inclusion of magnetized matter 
effects the two transverse DOF  $A_{\parallel}$, $A_{\perp}$ get coupled to 
each other, thus generating only {\it five} nonzero elements for the 
mixing matrix that is supposed to have nine elements . As one the 
includes the $\phi\gamma$  interaction, $A_{\parallel}$ gets coupled to 
$\phi$ and vice-versa, thus 
generating two more terms for the $3 \times 3$ mixing matrix, lifting the 
non-zero element count for the mixing matrix to {\it seven}. This can be 
verified from the explicit expression of the matrix. So the $3 \times 3$
mixing matrix has seven nonzero elements out of nine.\\

\noindent
{\bf Note}:  After this paper was submitted for publication, we came across the reference  \cite{palash2020}, that deals with some of the issues those are
 discussed in this work.





\begin{thebibliography}{99}


\bibitem{Kim} Y. M. Cho and J. H. Kim.
Phys. Rev. D {\bf 79}, 023504 (2009). \\

\bibitem{donoghue} Thibault Damour and John F. Donoghue,
Phys. Rev. D {\bf 82}, 084033 (2010). \\


\bibitem{gasperini} Maurizo Gasperini, 
\emph{ Elements Of String Cosmology}, Cambridge, UK, (Cambridge University Press 2007)\\.
\bibitem{bobby_Maharana} Bobby S. Acharya, Mansi Dhuria, Diptimoy Ghosh, Anshuman Maharana.
JCAP {\bf11}, 035 (2019).\\
\bibitem{coriano} Claudio Corian`, Luigi Delle Rose, Antonio Quintavalle and
Mirko Serino, JHEP, {\bf 077},1306 (2013); 
W. D.Goldberger, B. Grinstein and W. Skiba 
Phys. Rev. Lett. {\bf 100}, 111802 (2008). \\



\bibitem{salam} 
A. Salam and J.A. Strathdee, Phys. Rev. {\bf 184}, 1760 (1969).\\



\bibitem{fifthforce1} Y.M. Cho, Phys. Rev. D {\bf 41}, 2462 (1990).\\

\bibitem{fifthforce2} Y.M. Cho and J.H. Yoon, Phys. Rev. D {\bf 47}, 3465 (1993).\\
\bibitem{fifthforce3} Y.M. Cho, in Proceedings of XXth Yamada Conference, 
edited by S. Hayakawa and K. Sato (University Academy,Tokyo 1988).\\
\bibitem{fifthforce4} Koijam Manihar Singh, Kangujam Priyo kumar Singh,
Mod. Phys. Lett. A {\bf 34} 09, 1950056 (2019).\\
\bibitem{booby}Bobby Samir Acharya,  Anshuman Maharana, Francesco Muia. JHEP {\bf03}, 048 (2019).\\


\bibitem{bending}Dongjin Chway, 
arXiv:190109760 [gr-qc], kioon choi, Dongjin Chway, Chang 
Sub Shin,  JHEP {\bf 142}, 1811 (2018).\\
\bibitem{equiv} T. Damour and J. F. Donoghue, Phys.Rev. D {\bf 82}, 084033 (2010).  \\

 

\bibitem{Miani} L. Maiani, R. Petronzio and E. Zavattini.
Phys. Lett. B {\bf 175}, 359 (1986). \\


\bibitem{Raffelt}
G. Raffelt and L. Stodolsky. 
Phys. Rev. D {\bf 37}, 1237 (1988).\\



\bibitem{Kahniashvili} V. Baukh, A. Zhuk and T. Kahniashvili.
Phys. Rev. D {\bf 76}, 027502 (2007). \\






\bibitem{Ganguly-parthasarathy} A.K. Ganguly and R. Parthasarathy. 
Phys. Rev. D {\bf 68}, 106005 (2003).\\

 
 

\bibitem{Giannotti2} M. Giannotti, I. Irastorza, J. Redondo and A. Ringwald.
JCAP {\bf 05}, 057 (2016).\\



\bibitem{Ganguly-jaiswal} A. K. Ganguly and M. K. Jaiswal. 
Phys. Rev. D {\bf  90}, 026002 (2014).\\




\bibitem{shahbad} H. Perez Rojas and A. E. Shabad.  
Ann. Phys. (N.Y.) {\bf 121}, 432-455 (1979) \\




\bibitem{nieves} J. C.  DOlivo, J. F. Nieves and S. Sahu.
 Phys. Rev. D. {\bf 67} 025018, (2003).\\





\bibitem{servant}  J.R. Espinosa , C. Grojean, G. Panico, A. Pomarol, O. Pujolas, and G. Servant
Phys. Rev. Lett. {\bf 115}, 251803 (2015). \\
 





 
 
 
\bibitem{axion-fifthforce} S. A. Hoedl, F. Fleischer, E. G. Adelberger 
and B. R. Heckel.
Phys. Rev. Lett. {\bf 106}, 041801 (2011).\\





\bibitem{diff-axion-scalar-lab-tests} Sonny Mantry, Mario Pitschmann and 
Michael J. Ramsey-Musolf. 
arXiv:1411.2162v1.\\




\bibitem{Parthamajumdar} P. Majumdar.
Mod Phys. Lett. A {\bf19}, 1319(2004).\\




\bibitem{RaffeltChi} G.G. Raffelt.
Chicago Univ. Pr., Chicago U.S.A.(1996).\\



\bibitem{Ringwald} A. Ringwald. 
PoS(NEUTEL2015)021 arXiv:1506.04259. \\



\bibitem{Giannotti1} M. Giannotti. 
 arXiv:1508.07576. \\ 




\bibitem{ayla} A. Ayala, I. Dominguez, M. Giannotti, A. Mirizzi 
and O. Straniero.
Phys. Rev. Lett. {\bf 113}, 191302 (2014).\\





\bibitem{yanagida} Ken'ichi Saikawa and Tsutomu T. Yanagida. 
 arXiv:1907.07662v1. \\ 

\bibitem{Giannotti} M. Giannotti, I. G. Irastorza, J. Redondo,
A. Ringwaldd and K. Saikawad. 
JCAP {\bf 10}, 010 (2017). \\

\bibitem{Mclerran} Larry McLerran. J. Exp. Theor. Phys. {\bf120} 3, 376 (2015). \\

\bibitem{palash-prima}R. Chanda, J. F. Nieves and P. B. Pal.
Phys.Rev. D {\bf 37}, 2714 (1988).\\


\bibitem{Cast} K. Zioutas et al. (CAST Collaboration).
Phys. Rev. Lett. {\bf 94},121301 (2005).\\

\bibitem{Castnature} K. Zioutas et al. (CAST Collaboration).
Nat. Phys. {\bf 13}, 584 (2017).\\

\bibitem{Armengaud} E. Armengaud et al. 
 JINST 9 T05002. arXiv:1401.3233 (2014). \\
 
\bibitem{xenon} XENON100 collaboration, E. Aprile et al.
Phys. Rev. D {\bf  90}, 062009 (2014), [Erratum ibid.] {\bf D 95}, 029904 (2017)
arXiv:1404.1455.\\



\bibitem{MADMAX} A. Caldwell, G. Dvali, B. Majorovits, A. Millar, G. Raffelt, J. Redondo, O. Reimann, F. Simon, F. Steffen (MADMAX Working Group),
Phys. Rev. Lett. {\bf 118},  091801 (2017).\\ 



 \bibitem{DARWIN} DARWIN collaboration, J. Aalbers et al.
 JCAP {\bf 11}, 017 (2016).\\
 
\bibitem{Irastorza}  I. G. Irastorza, J. Redondo.
arXiv:1801.08127.\\

\bibitem{Millar}A. J. Millar, J. Redondo and F. D. Steffen. 
JCAP {\bf 10}, 006 (2017) \\



\bibitem{Kartavtsev} A. Kartavtsev, G. Raffelt, H. Vogel.
JCAP {\bf 024}, 01, 1701 (2017).\\




\bibitem{subir}  S. Sarkar.
 Rep.~Prog.~Phys. {\bf 59}, 12, 1493 (1996).\\


 
\bibitem{Masaki} Emi Masaki, Arata Aoki, Jiro Soda.
 Phys. Rev. D {\bf 96},  043519 (2017).\\


\bibitem{JKPS}  Avijit K. Ganguly, Manoj K. Jaiswal. 
J. Korean Phys.Soc. {\bf 72}, 1, 6 (2018).\\
  


 
\bibitem{manoj}  A. K. Ganguly, Ankur Chaubey, M. K. Jaiswal. 
Photon-Scalar 
Oscillation with Dim-5 operators: In a magnetizd media 
(to be communicated). \\
 


\bibitem{kolb-turner}  E. W. Kolb, M. S. Turner, T. P. Walker 
Phys. Rev. D {\bf 34}, 2197 (1986).\\
 



\bibitem{linde} A. D. Linde, Rep. Prog. Phys. 42, 389 (1979)\\
\bibitem{Fradkin} E. s. Fradkin, Proc. Lebedev Phys. Inst. (1965)(English trans. 1967 by Consultants Bureau, New York).\\
\bibitem{pal-david} Jose F. Nieves, Palash B. Pal and David G. Unger, Phys.Rev.D {\bf 28} 908 (1983).\\ 
\bibitem{bellac}Michel Le Bellac, Thermal field theory, Cambridge Monographs on Mathematical Physics.  \\





\bibitem{Hanson} A. Hanson, T. Regge, C. Teitelboim, Academia Nazionale dei Lincei, Rome (1976).\\
\bibitem{Dirac} P.A.M Dirac, Lectures on Quantum mechanics, Yeshiva University Press, New York, 1964.\\
\bibitem{ECG-MUKUNDA} E.C.G Sudarshan and N. Mukunda, John Wiley, New York (1974).\\
\bibitem{Maharana} J. Maharana, Pramana, {\bf 38}, 5 (1992).\\



%
 
 
 
\bibitem{cherkas}  S. L. Cherkas, K. G. Batrakov and D. Matsukevich  Phys. Rev. D {\bf 66}, 065011 (2002). \\


 
 

\bibitem{jack} J.D jackson. Third edition, {\it Classical Electrodynamics}, Wiley, (1998).\\ 
\bibitem{Itzykson-Zuber}  C. Itzykson and J. B. Zuber, {\it Quantum Field theory}, 
Tata Mc Graw Hill (London), (1980).\\



\bibitem{Adler} S. Adler. 
Ann. Phys. {\bf 67}, 599 (1971).\\ 


\bibitem{Ganguly-jain} A. K. Ganguly, P. Jain, and S. Mandal. 
Phys. Rev. D  {\bf 79}, 115014 (2009).\\





\bibitem{Raffelt-magnetized} N.V. Mikheev, G. Raffelt and L. A. Vassilevskaya 
Phys. Rev.D {\bf 58}, 055008 (1998).\\


\bibitem{Pankaj-Ralston} R. Das, P.Jain and J. P. Ralston,  R. Saha.
Pramana {\bf 70}, 439 (2008).\\
\bibitem{Ganguly-ann} A. K. Ganguly.
Ann. Phys. {\bf 321}, 6, 1457 (2006).\\


\bibitem{Tercas} H. Tercas, J. D. Rodrigues and J. T. Mendonca.
Phys. Rev. Lett. {\bf 120}, 181803 (2018).\\



\bibitem{subir-axion} F. Miniati, G. Gregori, B. Reville and S. Sarkar. 
Phys. Rev. Lett. {\bf 121}, 021301 (2018). \\


 
\bibitem{choi-kim} K. Choi, H. Kim and T. Sekiguchi
Phys. Rev. Lett. {\bf 121}, 031102 (2018). \\




  

\bibitem{ganguly-konar-pal} A. K. Ganguly, S. Konar and P.B. Pal.
 Phys. Rev. D {\bf 60}, 105014 (1999).\\

\bibitem{packzynski} B. Paczynski, ApJ. {\bf 308}, L43 (1986).\\


\bibitem{ellis} J. R. Ellis, N. E. Mavromatos, D.V. Nanopoulos and 
A. S. Sakharov, Astron. Astrophys {\bf 402}, 409 (2003).\\

\bibitem{rubbia} A. Rubbia  and A. Sakharov, Astroparticle Physics {\bf 29}, 20 
(2008).\\

\bibitem{yonetokuI} D. Yonetoku et. al.  ApJ, {\bf 758}, L1 (2012).\\



\bibitem{yonetokuII} K. Toma, S. Mukohyama, D. Yonetoku, T. Murakami, S. Gunji,  T. Mihara,
Y. Morihara, T. Sakashita, T. Takahashi, Y. Wakashima, H. Yonemochi, and N. Toukairin             
Phys. Rev. Lett. {\bf 109}, 241104 (2012).\\





\bibitem{waxman} Eli Waxman 
 Phys. Rev. Lett. {\bf 75}, 3 (1995).\\



\bibitem{RHESSI} \emph{https://hesperia.gsfc.nasa.gov/rhessi3/}. \\


\bibitem{SWIFT} \emph{https://www.nasa.gov/mission\_pages/swift/main}\\
	
\bibitem{wilczek}
Matthew Lawson, Alexander J. Millar, Matteo Pancaldi, 
Edoardo Vitagliano, Frank Wilczek, Phys. Rev. Lett. {\bf 123}, 14, 141802 
(2019).\\

\bibitem{Granot03} Granot, J., ApJ, {\bf596}, L17 (2003).\\
\bibitem{Lyutikov}Lyutikov, M., Pariev, V. I., $\&$  Blandford, R., ApJ, {\bf 597}, 998 (2003).\\
\bibitem{Granot12} Granot, J. Komissarov, S, S., $\&$ Spitkovsky, A., MNRAS, {\bf 411},1323 (2012). \\
\bibitem{Zhang} Zhang, B., $\&$ Yan, H., ApJ, {\bf 726}, 90 (2011).\\
\bibitem{Inoue}Inoue, T., Asano, K., $\&$ Ioka, K., ApJ, {\bf 734}, 77 (2011).\\






\bibitem{APEX}
E. Silver , H. Schnopper. 
Ronaldo Belazzini, Enrico Costa,  Giorgio Matt and Gianpiero Tagliaferri.
Cambridge University Press, (2010).\\



\bibitem{IXPE} M. C. Weisskopf   et al. 
RINP {\bf 6}, 1179 (2016).\\



\bibitem{Hooman} Hooman Davoudiasl, Patrick Huber. Phys. Rev. Lett. {\bf 97},141302 (2006).\\


\bibitem{ASTRO-2020}B. Rani et al. 
arXiv:1903.04607v1.
Z. Wadiasingh et al.
arXiv:1903.05648.\\

\bibitem{e-ASTROGAM}
V. Tatischeff  et al.  
arXiv:1706.07031v2.\\




\bibitem{palash-amj} J. F. Nieves and P. B. Pal.
 Am. J. Phys. {\bf 62}, 3 (1992).\\

\bibitem{palash2020} Palash B. Pal, Phys. Rev. D {\bf 102}, 036004 (2020). arXiv:2005.09376





\end{thebibliography}
\end{document}